\documentclass[preprint,10pt]{elsarticle}
\usepackage{booktabs}  
\usepackage{placeins} 
\usepackage{booktabs}
\usepackage{siunitx}

\usepackage{siunitx}   
\usepackage[T1]{fontenc}
\usepackage[utf8]{inputenc} 
\usepackage{amsmath,amssymb,mathtools}
\usepackage{hyperref}
\usepackage{bookmark} 
\usepackage{graphicx}
\usepackage{subfigure}
\usepackage{cleveref}

\journal{Physics of the Dark Universe}

\begin{document}

\begin{frontmatter}



\title{WIMP Freeze-out dynamics under Tsallis statistics }

\author[xd]{Matias P. Gonzalez}
\author[xd]{Roberto A. Lineros}
\affiliation[xd]{organization={Departamento de Física, Universidad Católica del Norte},
            addressline={Avenida Angamos~0610}, 
            city={Antofagasta},
            postcode={1240000}, 
            country={Chile}}

\begin{abstract}
We generalize thermal WIMP (Weakly Interacting Massive Particle) freeze-out within Tsallis nonextensive statistics. Using Curado-Tsallis $q$-distributions $f_q(E;\mu,T)$ we compute $q$-deformed number and energy densities, pressure, entropy density and Hubble rate, $\{n_q,\rho_q,P_q,s_q,H_q\}$. The Boltzmann equation is generalized accordingly to obtain the comoving abundance $Y_{\chi,q}(x)$ and relic density $\Omega_{\chi,q}h^2$ for a dark-matter candidate $\chi$ in a model-independent setup. The thermally averaged cross section is expanded as $\langle\sigma v\rangle_q \approx a + b\,\langle v_{\rm rel}^2\rangle_q$ up to $p$-wave. The freeze-out parameter $x_f(q)$ is determined from $\Gamma_{{\rm ann},q}(T_f)\simeq H_q(T_f)$ using a $q$-logarithmic inversion, with the expansion rate modified through ultra-relativistic rescalings $R_\rho(q)$ of the effective relativistic degrees of freedom $g_*$ and $g_{*s}$. We show that $x_f$ increases with $q$ and that QCD-threshold features propagate into $Y_{\chi,q}(x)$ and $\Omega_{\chi,q}h^2$. We then perform two $q$-grid scans:  fixing $\langle\sigma v\rangle_q$ while varying the dark-matter mass $m_\chi$, and fixing $m_\chi$ while varying the $s$-wave coefficient $a$. For an $s$-wave dominated scenario we construct $\chi^2$ profiles in these planes by comparing $\Omega_{\chi,q}h^2$ with the Planck benchmark $\Omega_c h^2 = 0.120\pm 0.001$. In both cases we find a clear degeneracy in the preferred nonextensive parameter $q_{\rm best}$ along valleys in parameter space. However, fixed-mass scans (varying $\langle\sigma v\rangle_q$) are significantly more constraining than fixed-cross-section scans, reflecting that $\Omega_{\chi,q}h^2$ is mainly controlled by $\langle\sigma v\rangle_q$, so that for realistic cross sections the best-fit $q_{\rm best}$ remains close to the extensive limit $q\to 1$.

\end{abstract}


\end{frontmatter}



\section{Introduction}
The standard cosmological model ($\Lambda$CDM) performs remarkably well: with a small set of parameters it provides a coherent description of observations ranging from CMB anisotropies to large-scale structure and the late-time expansion. However, this empirical success coexists with limitations that become relevant when connecting cosmology to early-Universe microphysics: the nature of dark matter and dark energy remains unknown, the origin of the required initial conditions is not addressed within the model, and it is not guaranteed that cosmological inferences remain unchanged if the primordial plasma departs, even mildly, from the strict equilibrium assumptions usually adopted. These considerations motivate exploring controlled extensions of the standard thermal picture and quantifying their impact on decoupling dynamics and derived cosmological observables \cite{2020}.

The existence of a dark matter component is one of the clearest indications of physics beyond the Standard Model of particle physics. In the early Universe, the primordial plasma was hot and dense enough that microscopic reactions efficiently created and destroyed particle species, tying their abundances which are determined by thermal decoupling (freeze-out) to the thermodynamic state of the bath. This makes the thermal history of the Universe a optimal scenario to connect statistical mechanics, particle physics and cosmological observations.

Thermal freeze-out links microphysics to cosmology: during radiation dominated era, frequent interactions keep dark-matter candidates in equilibrium with the primordial plasma until expansion outpaces reactions and the species decouple, leaving a nearly constant comoving abundance \cite{Scherrer:1986}. This paradigm has guided decades of work on WIMPs and related scenarios, where predictions are confronted with the observed cold dark matter density, $\Omega_{c}h^{2}= 0.120\pm 0.001$ \cite{2020}.

In the standard framework, we model freeze-out with a kinetic evolution for the comoving number density under radiation domination, using thermal averages of the annihilation rate and tracking thresholds in the effective degrees of freedom \cite{LaineSchroder:2006}. The literature refines this baseline with velocity-dependent annihilation, coannihilations, resonances, and nontrivial thermal histories, translating the evolution into constraints by comparison with $\Omega_{c}h^{2}$ \cite{GriestSeckel:1991,GondoloGelmini1991}.

In parallel, the particle-physics implementation of thermal dark matter has entered a strongly constrained stage. The absence of signals in the increasingly complementary program of direct detection, indirect searches, and collider probes has placed severe pressure on large classes of simple WIMP realizations, pushing many minimal SM-portal constructions into narrow and fine-tuned corners of parameter space \cite{Arcadi:2017kky}. In this setting, it becomes timely to reassess how robust the standard freeze-out picture is with respect to controlled deformations of its underlying assumptions, particularly those tied to the equilibrium statistical description of the radiation bath.

Moreover, within a nonextensive setup the mapping between microphysical parameters and cosmological observables can be quantitatively modified. Since the relic abundance inferred from freeze-out depends on both the expansion history and the thermal averages entering the annihilation rate, a controlled deformation can shift the required thermally averaged cross section $\langle\sigma v\rangle$ (and the corresponding couplings and masses) needed to reproduce the relic abundance $\Omega_{c}h^{2}$. As a result, regions of parameter space that appear excluded or highly tuned within the standard framework may be partially reopened once nonextensive effects in the radiation bath are consistently accounted for, offering a motivated way to reassess the viability of constrained WIMP-like scenarios \cite{Arcadi:2017kky}.

Alongside this setup, nonextensive statistical mechanics provides a controlled deformation of equilibrium weights through a single real parameter $q$ (where $q=1$ means the standard framework), motivated by a medium with long-range correlations, memory, or anomalous transport-features relevant to high-energy plasmas \cite{Tsallis1988}. In early-Universe applications, we can implement the deformation coherently in the thermodynamic background (impacting the expansion history) and in the kinetic description, with complementary strands in the literature \cite{LimaPlastino:2001}. And applications in the near extensive regime $q\simeq 1$ \cite{Pessah_2001}.

The literature has previously explored the implications of Tsallis statistics on dark matter cosmology.
Notably, works such as Rueter, Rizzo, and Hewett~\cite{rueter2020darkmatterfreezetsallis} have analyzed WIMP freeze-out by investigating an approach where the collision term in the Boltzmann equation is generalized from the first principles of entropy production.
Such an approach, while fundamental, leads to complex and non-factorizable collision integrals that replace the thermally averaged annihilation rate.
Furthermore, those models often rely on a specific parameterization of the evolution of $q$ (e.g., relaxing from $q_0>1$ to $q=1$ at a defined cutoff).

In contrast, the approach in this work is deliberately "model-independent" and phenomenological.
Instead of re-deriving the collision term, we preserve the standard partial-wave structure $\langle \sigma v \rangle_q \approx a + b \langle v_{\rm rel}^2 \rangle_q$, which maintains a direct connection to observable annihilation parameters.
We coherently introduce non-extensivity only in the components affected by the plasma's statistical mechanics:
in the thermal average $\langle v_{\rm rel}^2 \rangle_q$, calculated rigorously from the $q$-distributions, and
in the expansion rate $H_q$ and entropy density $s_q$ via a rescaling of the relativistic background.
This method allows us to quantify the impact of $q$ directly, avoiding assumptions about the form of the collision term or the relaxation history of $q$.

A recent Tsallis-based cosmology was proposed in ~\cite{e25111495}, where Jizba and Lambiase formulate the first two laws of thermodynamics for gravitating systems using Tsallis extensive but non-additive $\delta$-entropy and apply the resulting modified dynamics to a radiation-dominated universe. By confronting their framework with Big Bang nucleosynthesis and cold dark matter relic density, they show that a nearly extensive value $\delta\simeq 1.499$ (anomalous dimension $\Delta\simeq 0.0013$) can simultaneously accommodate light-element abundances and the observed dark matter density, providing a useful benchmark for Tsallis-inspired cosmological scenarios.


 We fix notation and minimal conventions, construct a consistent mapping that transports the $q$-deformation to the radiation bath, formulate and solve the kinetic evolution of the comoving abundance and extract the decoupling point and relic abundance. We then profile over $q$ and the $s$-wave dominated scenario to quantify shifts in freeze-out and identify degeneracies, benchmarking all predictions directly against $\Omega_{c}h^{2}$.

The rest of this paper is organized as follows. In Sec.~\ref{statistical:framework} we introduce the Tsallis nonextensive framework, define the $q$-distribution functions, and construct the associated cosmological observables. In Sec.~\ref{qgeneralizedboltzmann:sec} we formulate the $q$-generalized Boltzmann equation, specify the thermally averaged cross section, and derive the freeze-out condition. Section~\ref{results:Sec} presents our numerical results for the comoving abundance, relic density, freeze-out parameter, and the statistical analysis of nonextensivity. We conclude in Sec.~\ref{sec:conclusions}, while additional technical material is collected in the appendices.

\section{Framework}
\label{statistical:framework}

Before introducing the specific definitions used throughout this work, we briefly summarize the intuition behind Tsallis nonextensive statistics for readers less familiar with the framework. Tsallis approach provides a one-parameter deformation of the usual Boltzmann-Gibbs equilibrium description, replacing exponential weights by $q$-exponentials and thereby allowing for controlled departures from extensivity. The parameter $q$ quantifies how strongly the system deviates from the standard equilibrium assumptions, and can effectively capture the presence of long-range correlations, memory effects, or anomalous transport that make the ordinary additive entropy description less adequate. In the limit $q\to 1$ the formalism smoothly reduces to the standard extensive case, so that nonextensivity can be treated as a consistent extension, rather than a different theory of equilibrium thermodynamics.

\subsection{Tsallis entropy}
Tsallis Non-additive entropy \(S_q\) generalizes Boltzmann-Gibbs (BG) entropy by replacing the ordinary logarithm with its \(q\)-deformed counterpart \cite{Tsallis1988}. In units with \(k_B=1\),
\begin{equation}
S_q \equiv \frac{1 - \sum_i p_i^{\,q}}{q-1}\,,\qquad q\in\mathbb{R}\,,
\end{equation}
which recovers the standard BG case as \(q\!\to\!1\). The \(q\)-logarithm and \(q\)-exponential we use are defined by
\begin{equation}
\ln_q f \equiv \frac{f^{\,1-q}-1}{1-q}\,,\qquad
e_q(x) \equiv \big[\,1+(1-q)x\,\big]^{\!\frac{1}{1-q}}\!\label{eq:qexp},
\end{equation}
again yielding \(\ln_1 f=\ln f\) and \(e_1(x)=e^x\), see~\ref{anexo:1} for more details about this functions.
A key property of $S_q$ is the \emph{pseudo-additivity}: for statistically independent subsystems \(A\) and \(B\),
\begin{equation}
S_q(A{+}B) = S_q(A)+S_q(B) + (1-q)\,S_q(A)\,S_q(B)\,,
\end{equation}
which reduces to strict additivity at \(q=1\). This allows us to capture long-range correlations and constraints typical of complex systems.

\subsection{$q$-distribution functions}

To extremize \(S_q\) under macroscopic constraints, we use the Curado-Tsallis (CT) scheme, i.e. not normalized \(q\)-averages for energy and particle number,
\(
\overline{E}=\sum_i p_i^{\,q}E_i,\;
\overline{N}=\sum_i p_i^{\,q}N_i
\),
together with the standard normalization of the probabilities \(\sum_i p_i=1\) \cite{CuradoTsallis1991, TsallisMendesPlastino1998}. This choice will lead to compact, numerically stable expressions for cosmological observables.

This leads to the mean occupation numbers
\begin{align}
f_{q}(E;\mu,T) \;&=\;
\frac{1}{\big[\,1+(q-1)\beta(E-\mu)\,\big]^{\!\frac{1}{q-1}}+\xi}
\;\\&=\;\frac{1}{e_q\!\big(\beta(E-\mu)\big)+\xi}\,,
\qquad \beta\equiv 1/T,
\label{eq:fq}
\end{align}
with Bose-Einstein (BE) \(\xi=-1\), Fermi-Dirac (FD) \(\xi=+1\) , and Maxwell-Boltzmann (MB) \(\xi=0\) \cite{Buyukilic1995,TsallisMendesPlastino1998,Tsallis2009Book}.
In the BG limit \(q\!\to\!1\) one recovers the standard BE/FD/MB laws.
Throughout we take \(\mu=0\) for the early-universe plasma and the WIMP sector while in chemical equilibrium.
The \(q\)-exponential in Eq.~\eqref{eq:fq} fixes the support: for \(q<1\), a finite cutoff \(E-\mu\le T/(1-q)\); for \(q\ge 1\), power-law tails.

\subsection{Cosmological observables in nonextensive statistical mechanics}
Having specified the $q$-generalized distribution functions, we now construct the macroscopic observables that enter the cosmological evolution. In particular, we define the number density, energy density, and pressure associated with a given species in the nonextensive framework.

We work with vanishing chemical potential $\mu=0$. The generalized macroscopic observables: number density $n_q$, energy density $\rho_q$, and pressure $P_q$ follow the traditional integrals in the phase-space,
\begin{align}
    n_q &= \frac{g}{(2\pi)^3}\!\int f_q(E,T)\, d^3p,\\
    \rho_q &= \frac{g}{(2\pi)^3}\!\int E(p)\, f_q(E,T)\, d^3p, \\
    P_q &= \frac{g}{(2\pi)^3}\! \int \frac{p^2}{3E(p)}\,f_q(E,T)\,d^3p,
\end{align}
where $g$ denotes the internal degrees of freedom of the particle species and $f_q$ are the $q$-distribution functions~\eqref{eq:fq}. Introducing the dimensionless variables \cite{KolbTurner1990}
\begin{equation}
x \equiv \frac{m}{T}, \qquad y \equiv \frac{p}{T}, \qquad \frac{E}{T}=\sqrt{y^2+x^2},
\label{eq:variables-adim-exact}
\end{equation}
the above expressions become
\begin{align}
n_q(T) &= \frac{g}{2\pi^2}\,T^3 \int_{0}^{y_{\max}} y^2\, f_q\!\big(\sqrt{y^2+x^2}\big)\,dy,
\label{eq:nq-exacto}\\[4pt]
\rho_q(T) &= \frac{g}{2\pi^2}\,T^4\!
\begin{aligned}[t]
&\int_{0}^{y_{\max}} y^2\,\sqrt{y^2+x^2}\,\\[-0.3ex]
&\qquad \times f_q\!\big(\sqrt{y^2+x^2}\big)\,dy,
\end{aligned}
\label{eq:rhoq-exacto}\\[4pt]
P_q(T) &= \frac{g}{6\pi^2}\,T^4\!
\begin{aligned}[t]
&\int_{0}^{y_{\max}} \frac{y^4}{\sqrt{y^2+x^2}}\,\\[-0.3ex]
&\qquad \times f_q\!\big(\sqrt{y^2+x^2}\big)\,dy,
\end{aligned}
\label{eq:Pq-exacto}
\end{align}
with the $q$-dependent support (set of values where the function is nonzero)
\begin{equation}
y_{\max}(x,q)=
\begin{cases}
\sqrt{\big(\tfrac{1}{1-q}\big)^{\!2} - x^2}, & q<1,\\[4pt]
\infty, & q\ge 1.
\end{cases}
\label{eq:ymax-exact}
\end{equation}

 Using equation of state $P_q=\omega\rho_q$ which is preserved in this framework one obtains the entropy density for the relativistic case $\omega = 1/3$
\begin{equation}
s_q(T) = \frac{\rho_q(T)+P_q(T)}{T}
      = \frac{4}{3T}\,\rho_q(T).
\label{eq:sq-exacto}
\end{equation}

\begin{figure}[!h]
    \centering
    \includegraphics[width=1\linewidth]{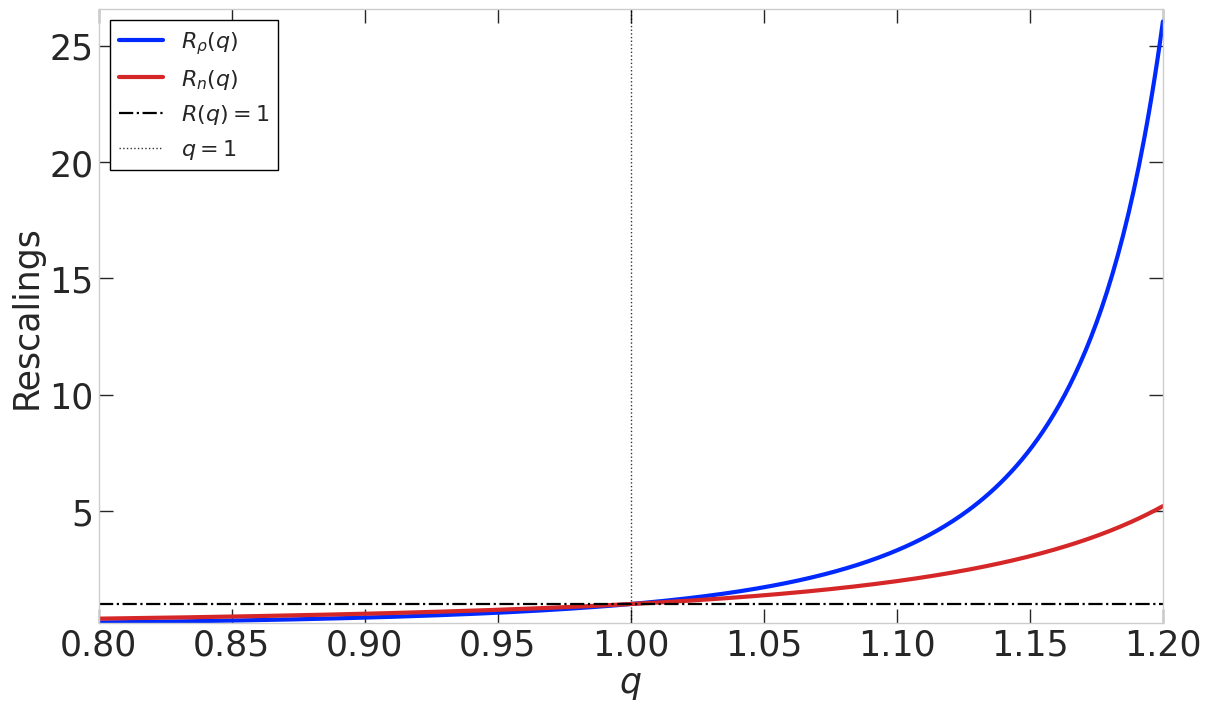}
   \caption{Nonextensive rescalings as functions of \(q\). Radiation rescaling \(R_\rho(q)\) is defined in Eq.~(\ref{eq:Rrho-exacto}) and computed as \(R_\rho=(1/6)\!\int_{0}^{z_{\max}} z^{3}\,e_q(-z)\,dz\), with \(z_{\max}=1/(1-q)\) for \(q<1\) and \(z_{\max}\!\to\!\infty\) for \(q\ge 1\). Convergence holds for \(q<5/4\). Analogously, the equilibrium number-density rescaling \(R_n(q)\), defined in Eq.~(\ref{eq:Rn-exacto}), is \(R_n=(1/2)\!\int_{0}^{z_{\max}} z^{2}\,e_q(-z)\,dz\) with the same \(z_{\max}\) prescription; its convergence condition is \(q<4/3\). Both mappings satisfy \(R_{\rho,n}(1)=1\) (dash-dotted line), and the dotted vertical line marks the extensive limit \(q=1\).}

    \label{fig:reescalados}
\end{figure}
In the ultra-relativistic (UR) limit ($E/T = p/T = z$) we introduce a rescaling map that relates nonextensive $q$ and extensive ($q\!\to\!1$) thermodynamic observables,
\begin{equation}
\rho_q(T)=R_\rho(q)\,\rho(T),\qquad
s_q(T)=R_\rho(q)\,s(T),
\label{eq:def-Rrho}
\end{equation}
where $R_\rho(q)$ encodes the effective $q$-rescaling of the thermal background. It is remarkable that we apply $R_{\rho}(q)$ in $\rho$ and $s$ because both describe the relativistic content in terms of degrees of freedom. Writing $z\equiv E/T$ and denoting by $e_q(x)$ the $q$-exponential, the factor $R_\rho$ reads
\begin{align}
R_\rho(q)\;\equiv\;
\frac{\displaystyle\int_{0}^{z_{\max}} z^{3}\,e_q(-z)\,dz}
     {\displaystyle\int_{0}^{\infty} z^{3}\,e^{-z}\,dz}
\;&=\;\frac{1}{6}\int_{0}^{z_{\max}} z^{3}\,e_q(-z)\,dz,
\label{eq:Rrho-exacto}
\end{align}
with
\begin{equation}
    z_{\max}(q)=\begin{cases}
\dfrac{1}{1-q}, & q<1,\\[4pt]
\infty, & q\ge 1,
\end{cases}
\end{equation}
and the integral converges for $q<5/4$ when $q\ge 1$ (since $e_q(-z)\sim z^{-1/(q-1)}$ and the integrand scales as $z^3$).

The rescaling \eqref{eq:def-Rrho} propagates into the effective and entropic relativistic degrees of freedom,
\begin{align}
g_{*,q}(T)\;=\;R_\rho(q)\,g_{*}(T),\label{eq:gstar_re}\\
g_{*s,q}(T)\;=\;R_\rho(q)\,g_{*s}(T),
\label{eq:gstarq-resc}
\end{align}
so that the cooling history uniformly rescales the ultra-relativistic plasma. The expansion rate in a flat Universe \cite{FRIEMAN_1994} is
\begin{equation}
    H\;=\;\sqrt{\frac{8\pi}{3}}\ \frac{\sqrt{\rho}}{M_{\rm Pl}},
\label{eq:H}
\end{equation}
hence in the nonextensive setting
\begin{equation}
    H_q(T)\;=\;\sqrt{\frac{8\pi}{3}}\ \frac{\sqrt{\rho_q(T)}}{M_{\rm Pl}}
\;=\; \sqrt{R_\rho(q)}\, H(T),
\label{eq:Hq_rescaled}
\end{equation}
which summarizes how the $q$-rescaling modifies both the thermodynamic background and the expansion history.

A fully analogous deformation follows from the number density,
\begin{equation}
n_q(T)=R_n(q)\,n(T),
\end{equation}
with the exact ultra-relativistic expression
\begin{align}
R_n(q)\;\equiv\;
\frac{\displaystyle\int_{0}^{z_{\max}} z^{2}\,e_q(-z)\,dz}
     {\displaystyle\int_{0}^{\infty} z^{2}\,e^{-z}\,dz}
\;=\;\frac{1}{2}\int_{0}^{z_{\max}} z^{2}\,e_q(-z)\,dz,
\label{eq:Rn-exacto}
\end{align}
using the same \(z_{\max}\) as the previous rescaling. For $q\ge 1$ this integral converges for $q<4/3$ (now the integrand scales as $z^2$), see~\ref{anexo:2} for convergence details. Both rescalings are compared in Fig. \ref{fig:reescalados}, is worth to mention that this rescalings are not dependent of the species (MB-statistics), we are making a global deformation for simplicity. 

\section{q-generalized Boltzmann equation for WIMP freeze-out}
\label{qgeneralizedboltzmann:sec}

\subsection{Generalizing the Boltzmann equation}
The starting point is the usual Boltzmann equation for the number density \cite{GondoloGelmini1991},
\begin{equation}
 	\frac{dn_\chi}{dt} + 3Hn_\chi = - \langle\sigma v\rangle \!\left( n_\chi^2 - n_{\chi,{\rm eq}}^{\,2} \right),
\end{equation}
with $\chi$ the WIMP candidate and $\langle\sigma v\rangle$ the thermally averaged cross section of annihilations. 
In a nonextensive medium we promote $n_\chi\!\to n_{\chi,q}$, defining the comoving abundance as $Y_{\chi,q}\!\equiv n_{\chi,q}/s_q$ and use $x\equiv m_\chi/T$. Using $s_q$ and $H_q$ from Eqs.~\eqref{eq:sq-exacto} and \eqref{eq:Hq_rescaled}, the $q$-generalized Boltzmann equation reads
\begin{align}
    &\frac{dY_{\chi,q}}{dx} = -\,\frac{s_q\,\langle\sigma v\rangle_q}{H_q\,x}
    \left(Y_{\chi,q}^{2}-Y_{\chi,q,{\rm eq}}^{2}\right) \nonumber\\
    &= -\,\sqrt{\frac{\pi}{45}}\;
       \frac{g_{*s,q}}{\sqrt{g_{*,q}}}\;
       m_\chi M_{\rm Pl}\,
       \frac{\langle\sigma v\rangle_q}{x^{2}}
       \left(Y_{\chi,q}^{2}-Y_{\chi,q,{\rm eq}}^{2}\right),
\label{eq:q_gen_boltzmann}
\end{align}
where $g_{*,q}$ and $g_{*s,q}$ are defined in Eqs.~\eqref{eq:gstar_re}-\eqref{eq:gstarq-resc} and $Y_{\chi,q,{\rm eq}} = n_{\chi,q,{\rm eq}}/s_q$ follows from Eqs.~\eqref{eq:nq-exacto}-\eqref{eq:Pq-exacto}. 
With the mapping $g_{*,q}=R_\rho(q)g_*$ and $g_{*s,q}=R_\rho(q)g_{*s}$ one may write the prefactor as
\begin{equation}
\frac{g_{*s,q}}{\sqrt{g_{*,q}}}=\sqrt{R_\rho(q)}\;\frac{g_{*s}}{\sqrt{g_*}},
\end{equation}
i.e. the canonical coefficient is rescaled by $\sqrt{R_\rho(q)}$.
Equation~\eqref{eq:q_gen_boltzmann} governs the freeze-out of the comoving abundance for $q\neq 1$; the only model-dependent input is the $q$-generalized thermal average $\langle\sigma v\rangle_q$, discussed next.

\subsection{On the thermally averaged cross section}
The thermally averaged annihilation rate provides the effective interaction strength in a hot plasma. For WIMP annihilation we adopt the standard partial-wave expansion up to $p$-wave \cite{GondoloGelmini1991},
\begin{equation}
\langle \sigma v \rangle \approx a + b\,\langle v_{\rm rel}^{2}\rangle,
\label{eq:ondasparcialesnormal}
\end{equation}
where $a$ and $b$ encode the $s$ and $p$-wave contributions respectively and $\langle v_{\rm rel}^{2}\rangle$ is the mean value of the relative velocity squared. We will be using $\mathrm{GeV}^{-2}$ for thermally averaged cross section units. In the nonextensive framework we generalize $\langle v_{\rm rel}^{2}\rangle$ to its $q$-generalized counterpart $\langle v_{\rm rel}^{2}\rangle\to\langle v_{\rm rel}^{2}\rangle_{q}$ while satisfying $\langle v_{\rm rel}^2\rangle_q = 2 \langle v ^2\rangle_q$ since the particles involved in annihilations are both described by the same statistical weight and are in equilibrium. Then, the $q$-averaged squared velocity is defined as:
\begin{equation}
\langle v^{2}\rangle_q=
\frac{%
\begin{aligned}[t]
&\displaystyle \int_{0}^{y_{\max}}\! dy\, y^{2}\,
\frac{y^{2}}{y^{2}+x^{2}}\,
e_q\!\left(-\sqrt{y^{2}+x^{2}}\right)
\end{aligned}
}{%
\begin{aligned}[t]
&\displaystyle \int_{0}^{y_{\max}}\! dy\, y^{2}\,
e_q\!\left(-\sqrt{y^{2}+x^{2}}\right)
\end{aligned}
},
\label{exactovelocidadcuadrada}
\end{equation}
with $x\equiv m_\chi/T$, $y\equiv p/T$, and $y_{\max}$ given in Eq.~\eqref{eq:ymax-exact}. 
The numerator has the single-particle nonrelativistic moment $v^2=p^2/E^2=y^2/(y^2+x^2)$ consistent with our $q$-distributions at $\mu=0$; in the BG limit $q\to 1$ one recovers the usual Maxwell-Boltzmann result. By using \eqref{exactovelocidadcuadrada} in \eqref{eq:ondasparcialesnormal} the $q$-generalized partial-wave expansion reads
\begin{align}
    \langle \sigma v \rangle_q \approx a + b\,\langle v_{\rm rel}^2 \rangle_q\,.
\label{eq:partialwavesqgem}
\end{align}
This definition preserves the $s{+}p$ structure and captures the $q$-dependent reshaping of the high-energy tails (for $q>1$) and the finite-support cutoff (for $q<1$) both of which impact $\langle v_{\rm rel}^{2}\rangle$ and hence the effective annihilation rate. The expression \eqref{exactovelocidadcuadrada} will be solved numerically for simplicity in the $q$-Boltzmann equation solution for $Y_{\chi,q}(x)$. We keep the same coefficients $(a,b)$ as phenomenological parameters. This defines the scope of our model-independent approach: we assume that the non-extensive deformation $q$ manifests in the statistical mechanics of the plasma (i.e., the phase-space distributions) but not in the underlying quantum field theory matrix elements that define the $a$ and $b$ coefficients themselves. A derivation of $q$-dependent microphysics is beyond the scope of this phenomenological framework. 

\subsection{Relic Abundance computation}
The relic abundance follows from $Y_{\infty,q}$ in the definition of density parameter for WIMP dark matter as follows:
\begin{equation}
    \Omega_{\chi,q} h^2 = \frac{\rho_{\chi,q}}{\rho_{c}}h^2 = \frac{m_{\chi}s_{0}Y_{\infty,q}h^2}{\rho_c}, \label{eq:omegadarkmatter}
\end{equation}
where $s_0 = 2.9 \times 10^3 ~\rm{cm}^{-3}$ is the present day entropy density, $\rho_c = 1.0537 \times 10^{-5}~h^2 ~\rm{GeV}~\rm{cm}^{-3}$ is the critical energy density for the universe to have flat space-time geometry \cite{2020} and $h$ is the dimensionless Hubble constant \cite{PDG2024_AstroConsts}. 

\subsection{Freeze-out location}

Starting from the freeze-out condition that equates the annihilation rate to the $q$-rescaled Hubble expansion at the decoupling temperature $T_f$,
\begin{equation}
  \Gamma_{\mathrm{ann},q}(T_f)\;\simeq\;H_q(T_f),
\end{equation}
and adopting the partial-wave ansatz Eq.~(\ref{eq:partialwavesqgem}) one obtains the following $q$-logarithmic transcendental equation for the freeze-out parameter $x_f\equiv m/T_f$:
\begin{align}
x_f(q) \;\simeq\;
\ln_q\!\Biggl[
\frac{g_{\chi}\,M_{\rm Pl}\,m_{\chi}}{1.66\,(2\pi)^{3/2}\,\sqrt{g_*(T_f)\,R_\rho(q)}}
\times \;\langle \sigma v \rangle_q\,\sqrt{x_f(q)}
\Biggr].
\label{eq:trascendent}
\end{align}
Here \(g_{\chi}\) is the internal degrees of freedom of the WIMP, \(m_{\chi}\) its mass, \(g_*(T_f)\) the effective relativistic degrees of freedom evaluated at \(T_f\), and \(R_\rho(q)\) the nonextensive radiation rescaling defined in Eq.~\eqref{eq:Rrho-exacto}. 

We do not introduce any number-density rescaling in either the freeze-out condition of Eq.~\eqref{eq:trascendent} or in the $q$-Boltzmann equation (cf. Eq.~\eqref{eq:q_gen_boltzmann}) because the simple map in Eq.~\eqref{eq:Rn-exacto} is obtained in the UR limit and is therefore not valid at chemical decoupling, where the WIMP is nonrelativistic. Using $R_n$ in this regime would lead to an unjustified deformation of the dark sector through number density. Instead, all quantities that enter the annihilation rate
$
\Gamma_{\mathrm{ann},q}\;\sim\; n_{\chi,{\rm eq},q}\,\langle\sigma v\rangle_q
$
and the collision term are computed from the exact phase-space integrals: the equilibrium density $n_{\chi,{\rm eq},q}$ from Eq.~\eqref{eq:nq-exacto} evaluated in the nonrelativistic regime, and the velocity moment $\langle v^2\rangle_q$ from Eq.~\eqref{exactovelocidadcuadrada}, which we use to construct $\langle\sigma v\rangle_q$ consistently. In short, we avoid any ad hoc rescaling of number density $n$ and rely exclusively on the exact $q$-deformed integrals to determine the dynamics. On the other side we use only $R_{\rho}(q)$ for the radiation content since it is in the ultra relativistic regime during the WIMP freeze-out.

\section{Results and Discussion on WIMP Freeze-out}
\label{results:Sec}
Having set up the $q$-generalized thermodynamic observables and the corresponding Boltzmann evolution, we now explore the numerical impact of nonextensivity on thermal freeze-out. We first show how the comoving abundance $Y_{\chi,q}(x)$ responds to variations in $q$ for fixed $(m_\chi,a,b)$, highlighting the role of power-law tails and $q$-dependent support in delaying or advancing decoupling. We then translate these effects into the relic abundance $\Omega_{\chi,q}h^2$ as a function of the WIMP mass, emphasizing the interplay between nonextensive statistics and the QCD crossover in the effective degrees of freedom. Finally, we analyze the behavior of the freeze-out parameter $x_f(q,m_\chi)$ and quantify the sensitivity to $q$ through simple $\chi^2$ scans in the $(q,m_\chi,a)$ space, illustrating the resulting degeneracies and the extent to which current relic-density measurements constrain departures from extensivity.
\subsection{Comoving Abundance}

\begin{figure}[!h]
    \centering
    \includegraphics[width=1\linewidth]{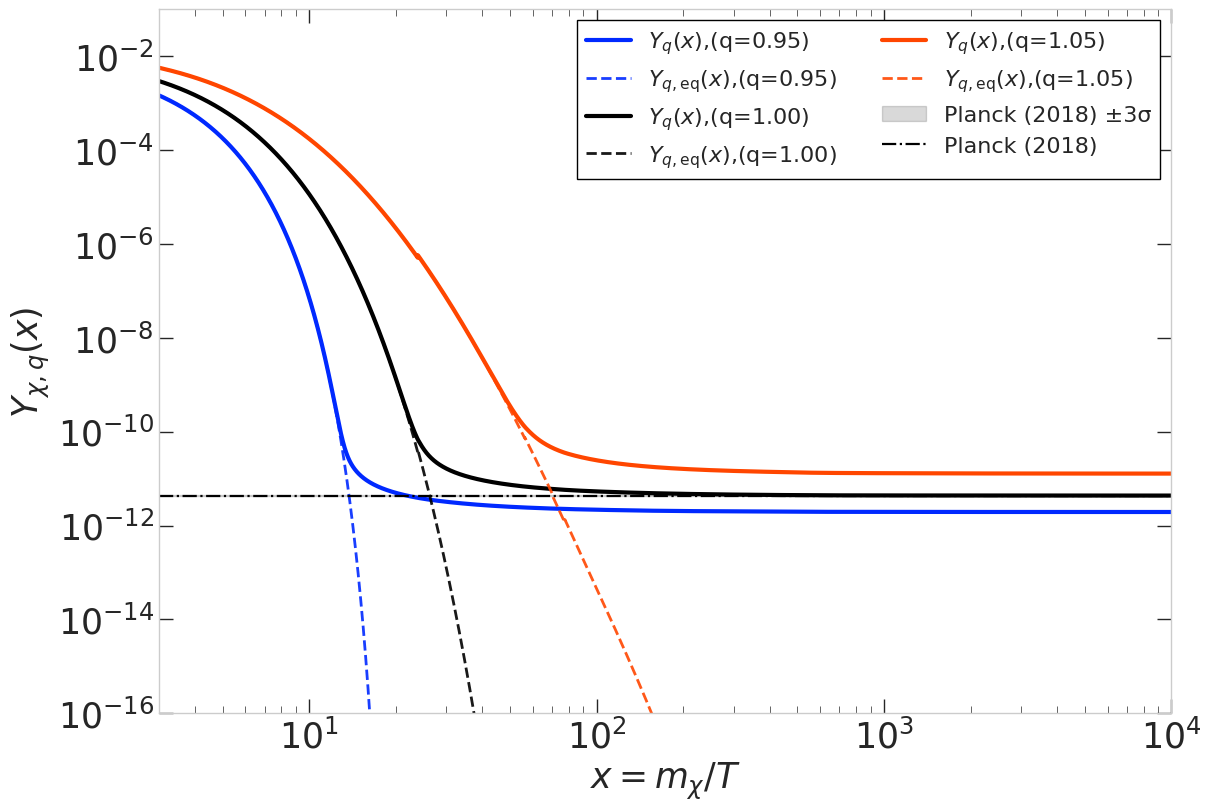}
    \caption{Freeze-out for $m_\chi=100~\mathrm{GeV}$, comoving abundance $Y_{\chi,q}(x)$ versus $x_f$: where $g_{\chi} = 4$, $a = 1.825 \times 10^{-9}~\rm GeV^{-2}$ and $b =  1.05 \times 10^{-9} ~\rm GeV^{-2}$ and several values of $q$ are considered. The black dashed line is the value measured by the Planck satellite $\Omega_c h^2 = 0.120 \pm 0.001$.}
    \label{fig:comovingabundance}
\end{figure}

                       As shown in Fig.~\ref{fig:comovingabundance}, the first panel (Yield vs.~$x$) highlights how nonextensivity reshapes the comoving abundance $Y_{\chi,q}(x)$ for fixed $m_\chi = 100 ~\rm{GeV}$. For $q>1$, the power-tails of the $q$-exponential enhance the velocity moments entering $\langle\sigma v\rangle_q\approx a+b\langle v_{\rm rel}^2\rangle_q$, so $Y_{\chi,q}(x)$ remains closer to equilibrium over a broader range in $x$. For $q<1$, the finite-support  (see Eq.~\eqref{eq:ymax-exact}) suppresses those moments, and $Y_{\chi,q}(x)$ departs from equilibrium more sharply. 

\subsection{Relic Abundance}

\begin{figure}[!h]
    \centering
    \includegraphics[width=1\linewidth]{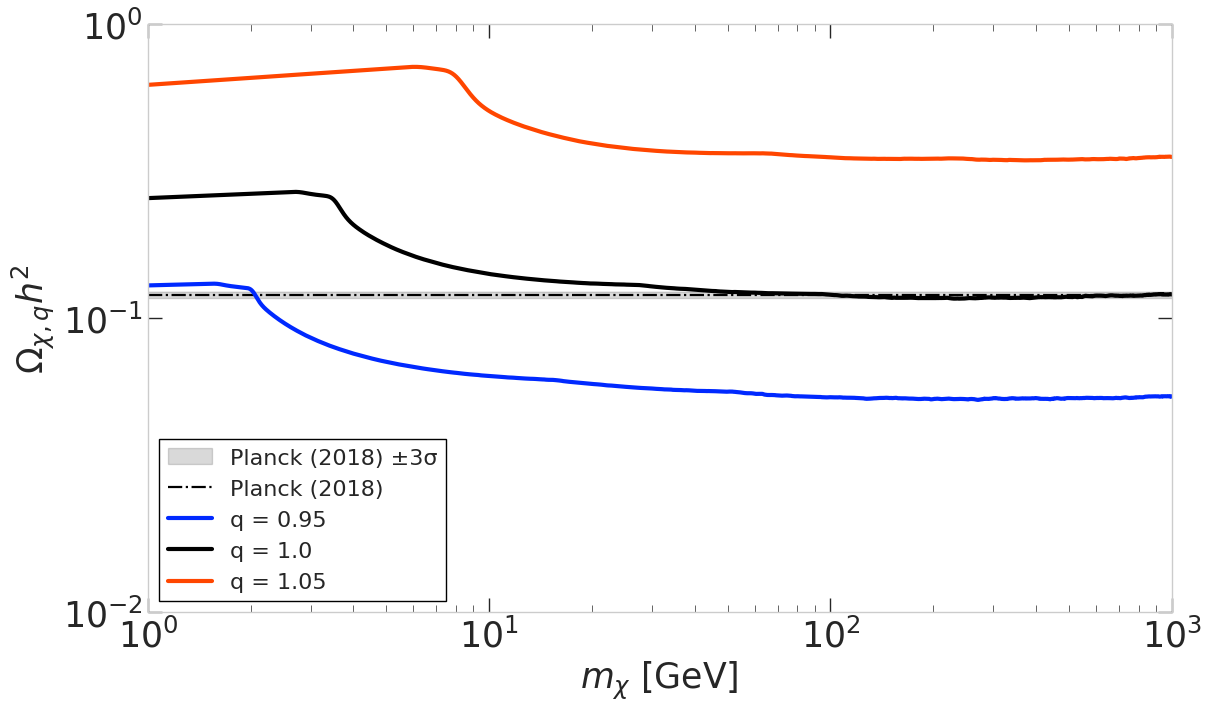}
    \caption{Relic abundance $\Omega_{\chi,q} h^2$ vs.\ mass $m_\chi$ for different $q$ for $g_{\chi}=4$.
  The shaded band and dashed line indicate, respectively, the $\pm3\sigma$ region and the central Planck value  $\Omega_c h^2 = 0.120 \pm 0.001$ with the same annihilation parameters as in Fig.~\ref{fig:comovingabundance}.}
    \label{fig:Relicabundance}
\end{figure}

In Fig.~\ref{fig:Relicabundance} the relic abundance \(\Omega_{\chi,q} h^2\) shows a marked drop when the mass \(m_\chi\) is such that the decoupling temperature \(T_f \simeq m_\chi/x_f\) enters the QCD crossover region (\(T \sim 150\text{-}170~\mathrm{MeV}\)), this can be seen where the relic abundance suddenly decreases. In that interval the plasma equation of state changes rapidly and the effective degrees of freedom \(g_{*,q}(T)\) and \(g_{*s,q}(T)\) decrease notably, altering both \(H_q \propto \sqrt{g_{*,q}}\,T^2\) and \(s_q \propto g_{*s,q}\,T^3\) in the Boltzmann equation. Since around Freeze-out one approximately has \(Y_{\infty,q} \sim H_q/\!\left(s_q\,\langle\sigma v\rangle_q\right)\), a reduction of \(g_{*,q}\) and \(g_{*s,q}\) translates into a visible decrease of \(\Omega_q h^2\). 

This effect is purely thermodynamic (intrinsic to the QCD plasma) and should not be confused with the opening or closing of annihilation channels \cite{Hindmarsh_2005,Saikawa_2020}. We also observe how the predicted relic abundance deviates from the measured one  $\Omega_c h^2 = 0.120 \pm 0.001$~\cite{2020} for $q \neq 1$, meaning that the standard annihilation parameters are not enough to reproduce what the experiments measured. 


\subsection{Freeze-out parameter}
\begin{figure}[!t]
    \centering
    \includegraphics[width=1\linewidth]{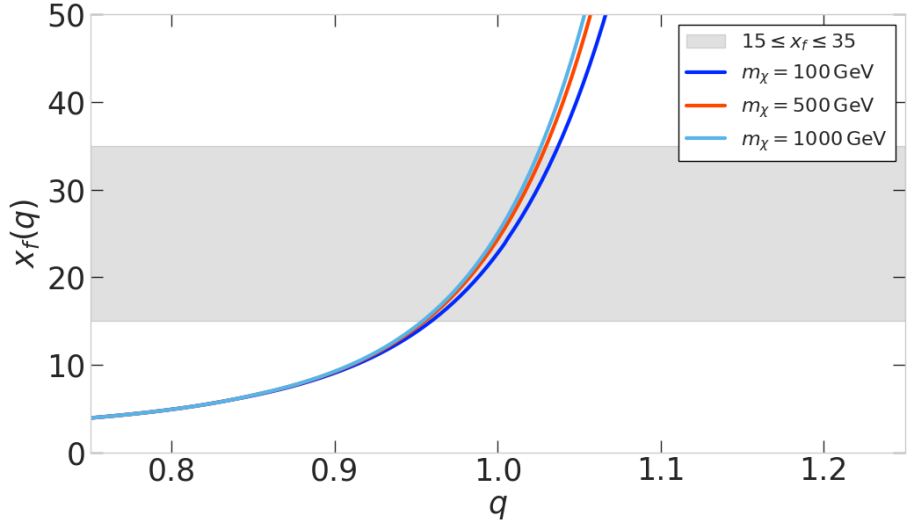}
    \caption{Freeze-out parameter $x_f$ versus nonextensivity $q$ for fixed WIMP masses
($m_\chi=\{100,500,1000\}\,\mathrm{GeV}$). Curves are obtained by solving the
$q$-generalized transcendental condition
Eq.~\eqref{eq:trascendent},
$R_\rho(q)$ computed from the exact $q$-exponential, and
$\langle\sigma v\rangle_q\approx a+b\,\langle v_{\rm rel}^2\rangle_q$.
The grey band indicates a ``canonical'' range $15\leq x_f \leq 35$.}
    \label{fig:trascendent}
\end{figure}

Figure~\ref{fig:trascendent}
shows $x_f(q)$ for $m_\chi=\{100,500,1000\}\,\mathrm{GeV}$ together with a canonical band
$15\le x_f\le 35$. For fixed $m_\chi$, $x_f$ increases monotonically with $q$: although
$H_q\!\propto\!\sqrt{R_\rho(q)}$ grows for increasing values of $q$ (which would tend to reduce $x_f$), the nonlinear
$q$-logarithmic mapping required to invert $e_q$ dominates in the range of interest, yielding a net
increase of $x_f$. This trend is essentially unchanged when setting $b=0$, confirming that the main
driver here is the $\ln_q$ inversion rather than the $p$-wave piece, see Eqs.~\eqref{eq:qexp}.

For \(q>1\), the exact \(q\)-exponential
\(e_q(-z)\) with \(z\equiv E/T\) decays as a power law Eq.~\eqref{eq:qexp},
enhancing the high-energy tail. This increases moments such as
\(\langle v_{\rm rel}^2\rangle_q\) entering the partial-wave approximation
\(\langle\sigma v\rangle_q\approx a+b\,\langle v_{\rm rel}^2\rangle_q\) Eq.~\eqref{eq:partialwavesqgem};
see also Eq.~\eqref{exactovelocidadcuadrada}.
The radiation background rescales as \(R_\rho(q)\) Eq.~\eqref{eq:Rrho-exacto},
so that \(g_{*,q}(T)=R_\rho(q)\,g_*(T)\) and
\(H_q(T)=\sqrt{R_\rho(q)}\,H(T)\) Eqs.~\eqref{eq:gstarq-resc}, \eqref{eq:Hq_rescaled}.
When the freeze-out condition is inverted with the \(q\)-logarithm
Eq.~\eqref{eq:trascendent}, this nonlinear mapping together with the larger
velocity moments yields a net delay of decoupling, hence a larger
\(x_f(q)\equiv m_\chi/T_f\). As \(q\to 5/4^{-}\), i.e. the convergence limit for $R_{\rho}(q)$ in the ultra relativistic limit, the
growth of \(x_f\) steepens.
\begin{figure}[!t]
    \centering
    \includegraphics[width=1\linewidth]{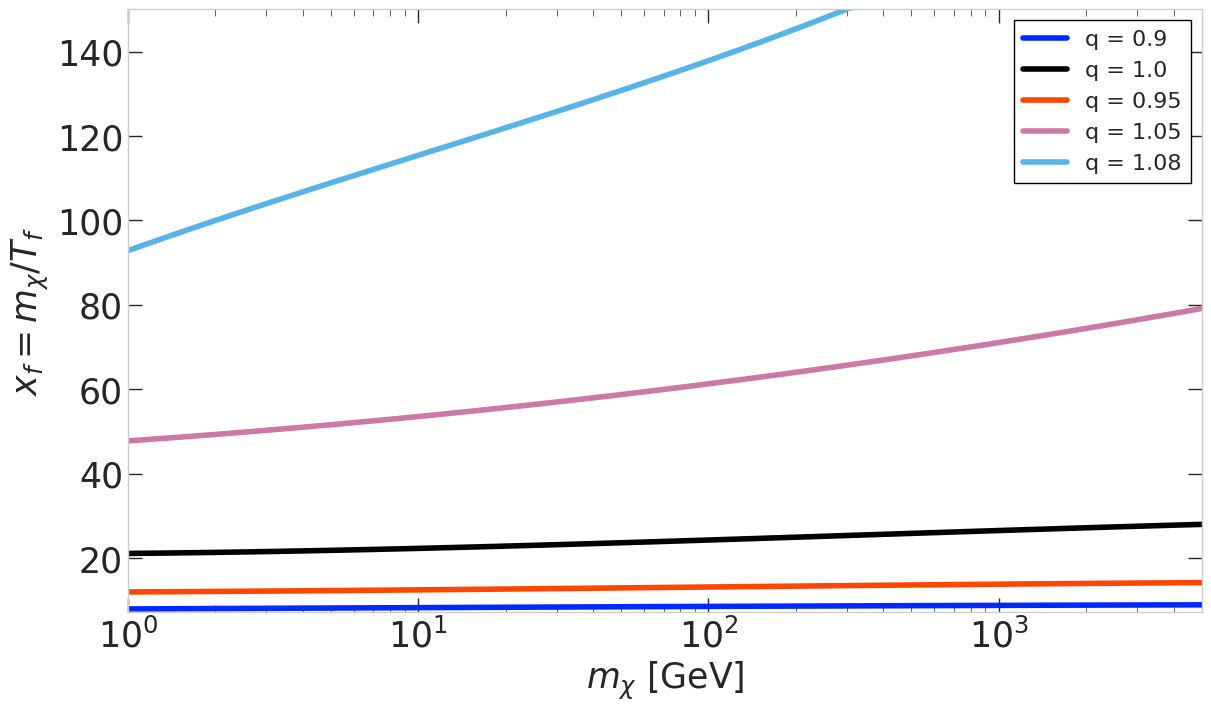}
   \caption{Freeze-out parameter \(x_f \equiv m_\chi/T_f\) as a function of \(m_\chi\) for \(q \in \{0.90,\,0.95,\,1.00,\,1.05,\,1.08\}\). Curves are obtained by numerically solving the \(q\)-generalized Boltzmann equation for \(Y_{\chi,q}\) with exact \(q\)-exponential distributions,  $\langle\sigma v\rangle_q\approx a+b\,\langle v_{\rm rel}^2\rangle_q$. The overall trend shows larger \(x_f\) for larger \(q\), consistent with the \(q\)-logarithmic inversion and the rescaled expansion rate \(H_q \propto \sqrt{R_\rho(q)}\).}
    \label{fig:xf_versus_m}
\end{figure}
For \(q<1\), \(e_q(-z)\) vanishes for
\(z\ge z_{\max}=1/(1-q)\) Eq.~\eqref{eq:qexp}, implying finite support in
energy/momentum as in Eq.~\eqref{eq:ymax-exact}. This suppresses
\(\langle v_{\rm rel}^2\rangle_q\), while the background is reduced
by \(R_\rho(q)<1\), giving \(H_q(T)=\sqrt{R_\rho(q)}\,H(T)<H(T)\)
Eqs.~\eqref{eq:gstarq-resc}, \eqref{eq:Hq_rescaled}. The suppression of the annihilation
rate \(\Gamma_{\mathrm{ann},q}\sim n_{\chi,{\mathrm{eq}},q}\,\langle\sigma v\rangle_q\) dominates over
the decrease in \(H_q\), so \(\Gamma_{\mathrm{ann},q}/H_q\) falls below unity at higher
temperatures and decoupling occurs earlier as $q$ decreases.

As shown in Fig.~\ref{fig:xf_versus_m}, for fixed $q$ the freeze-out parameter $x_f$ grows monotonically with the WIMP mass $m_\chi$. In the standard case ($q=1$) this is the usual logarithmic behaviour from $\Gamma_{\mathrm{ann}}\!\sim\!H$: as $m_\chi$ increases, the freeze-out temperature $T_f$ rises more slowly than $m_\chi$, so $x_f\equiv m_\chi/T_f$ increases steadily. Within the Tsallis framework ($q\neq1$) the trend is maintained but becomes $q$-dependent: both the slope and the offset of $x_f(m_\chi)$ are modified. The slope of $x_f(m_\chi)$ increases with $q$, with kinks where $g_{*,q}(T)$ and $g_{*s,q}(T)$ vary rapidly.

\subsection{Impact of nonextensivity}
\label{sec:sens}

To assess how nonextensivity impacts our results, we perform a simple statistical analysis against the observed relic abundance. In our model-independent setup, multiple combinations $(a,q,m_{\chi})$ which is the $s$-wave dominated scenario,  reproduce the Planck value $\Omega_{c}h^2 = 0.120\pm0.001$, so the constraint from a single observable is under-determined in the three-dimensional space. Consequently, the global minimum of a one effective parameter $\chi^2$ fit (built from the prediction $\Omega_{\chi,q}h^2$ computed by solving the $q$-generalized Boltzmann equation Eq.~\eqref{eq:q_gen_boltzmann} and the observed relic abundance $\Omega_{c}h^2 = 0.120\pm 0.001$) is not unique but organized along a valley (ridge) of nearly equivalent solutions.

To explore the parameter space in the simplest case of $s$-wave domination, two complementary scans are performed. At fixed annihilation cross section $\langle\sigma v\rangle_q$, $\chi^2(q)$ is evaluated on a grid in $q$ for a list of WIMP masses $m_\chi$. At fixed mass $m_\chi$, $\chi^2(q)$ is computed for a list of cross sections $\langle\sigma v\rangle_q$. These two slices make the $(q, m_\chi, a)$ landscape tractable, revealing the sensitivity of the relic abundance and the fit applied under parameter modifications.

\begin{figure}[!h]
    \centering
    \includegraphics[width=0.9\linewidth]{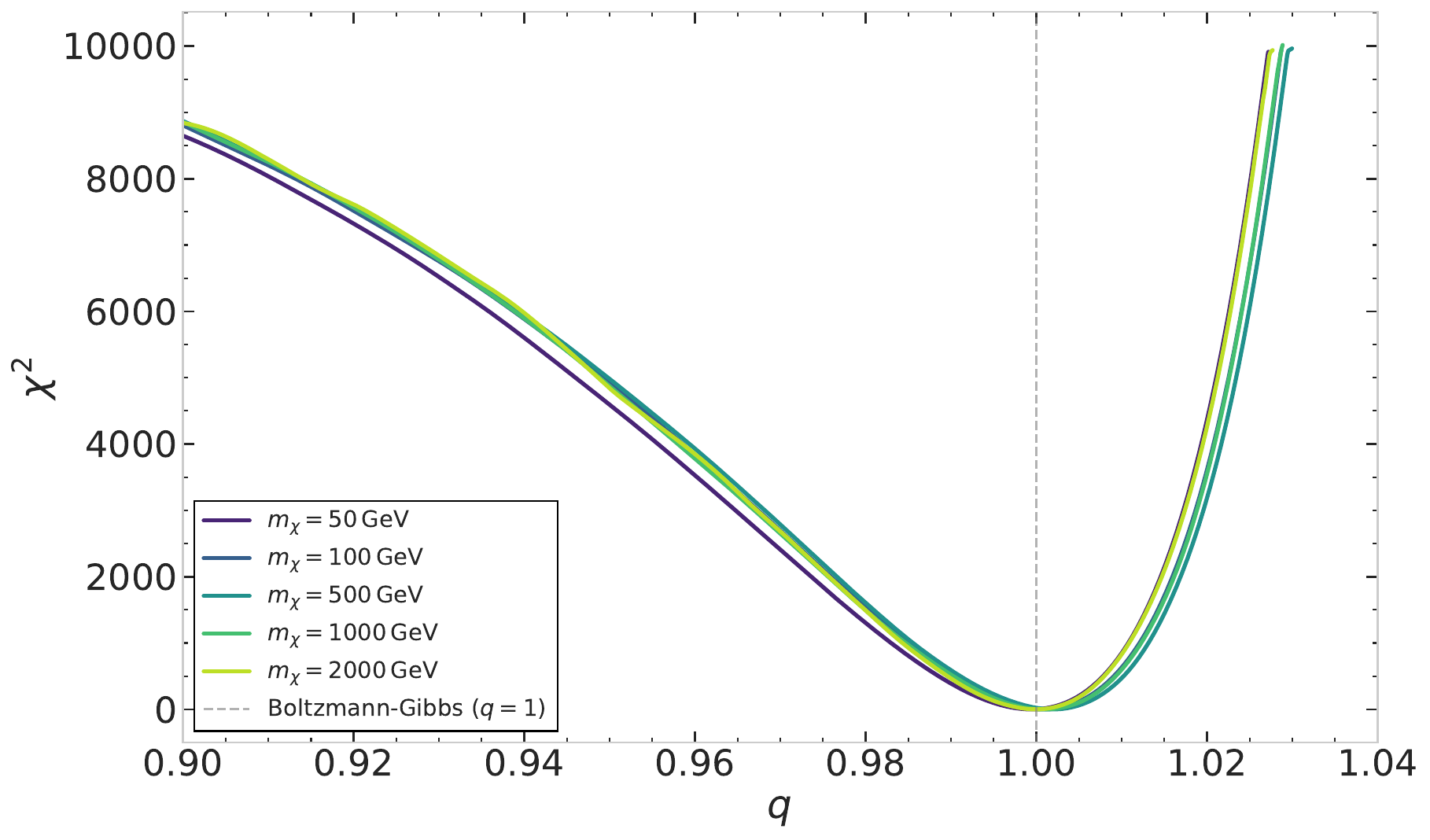}
    \caption{Absolute $\chi^2(q)$ profiles at fixed cross section (mass scan). Smoothed curves for several $m_\chi$ values illustrate the global fit structure and the approximate degeneracy in $q$ when only $\Omega_{\chi}h^2$ is used as constraint; the common minimum indicates nearly equivalent solutions across masses. A representative fixed thermally averaged cross section $\langle \sigma v \rangle_q = 2 \times 10 ^{-9}\mathrm{GeV}^{-2}$ was used.}
    \label{fig:chimasa}
\end{figure}

\begin{figure}[!h]
    \centering
    \includegraphics[width=0.9\linewidth]{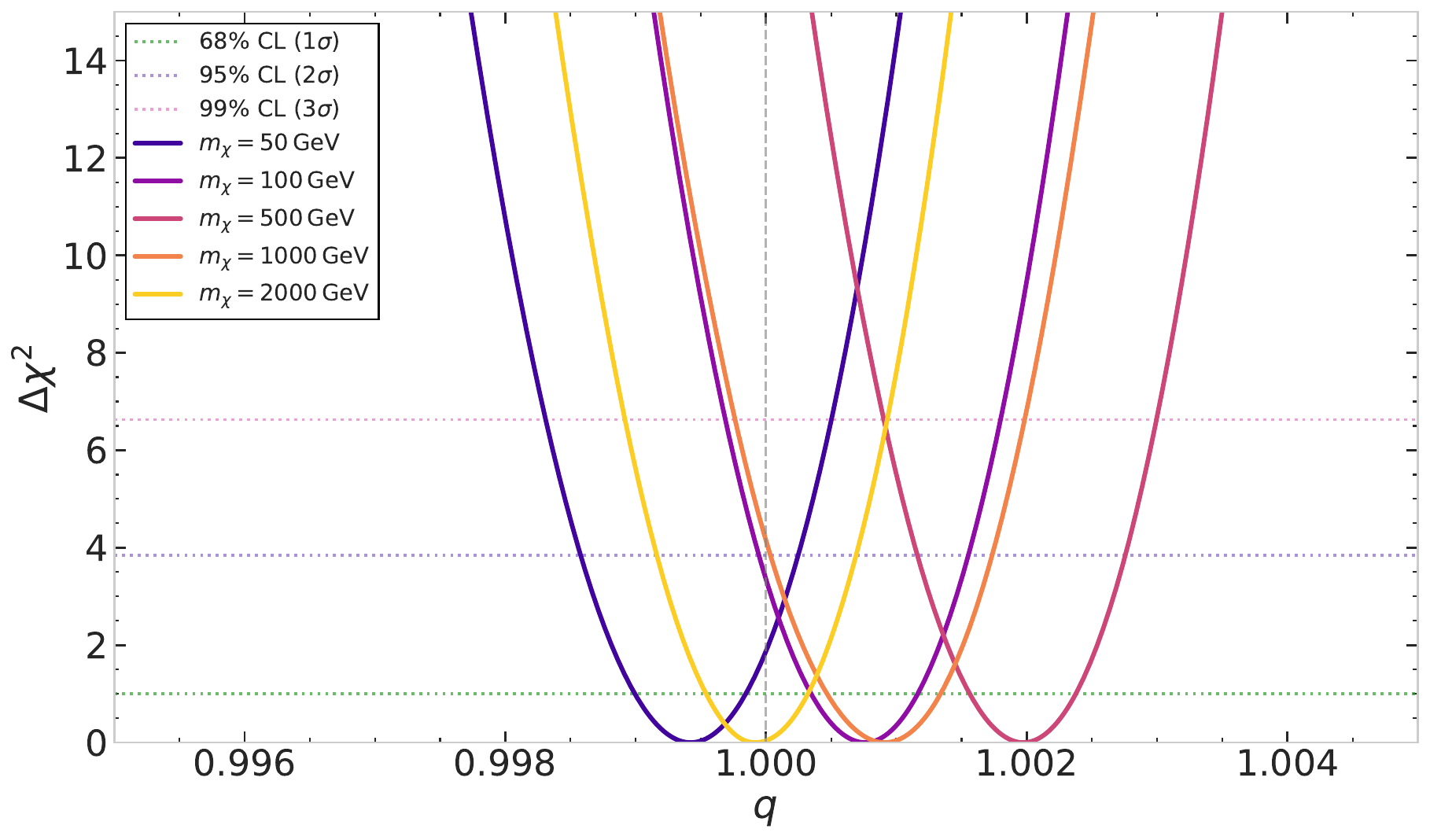}
        \caption{Profile likelihood $\Delta\chi^2(q)$ (mass scan). We show $\Delta\chi^2(q)=\chi^2(q)-\chi^2_{\min}$ for the same masses as above. Horizontal lines mark the 68\%, 95\%, and 99\% confidence levels for one effective parameter, defining the allowed $q$ intervals around $q_{\rm best}$.}
    \label{fig:chimasadelta}
\end{figure}

We begin with the mass scan at fixed thermally averaged cross section (Figs.~\ref{fig:chimasa} and \ref{fig:chimasadelta}). The absolute $\chi^2(q)$ and the profiled $\Delta\chi^2(q)$ show tightly clustered minima across $m_\chi$ values, and the $68\%\text{-}95\%$ confidence bands overlap broadly. This keeps the favored $q$ close to the extensive limit $q\simeq 1$. Physically, at fixed $a$ the mass mainly enters through the mild, logarithmic dependence of the freeze-out parameter $x_f$ on $m_\chi$, so changing $m_\chi$ barely shifts the location of the minimum.

\begin{figure}[!h]
    \centering
    \includegraphics[width=0.9\linewidth]{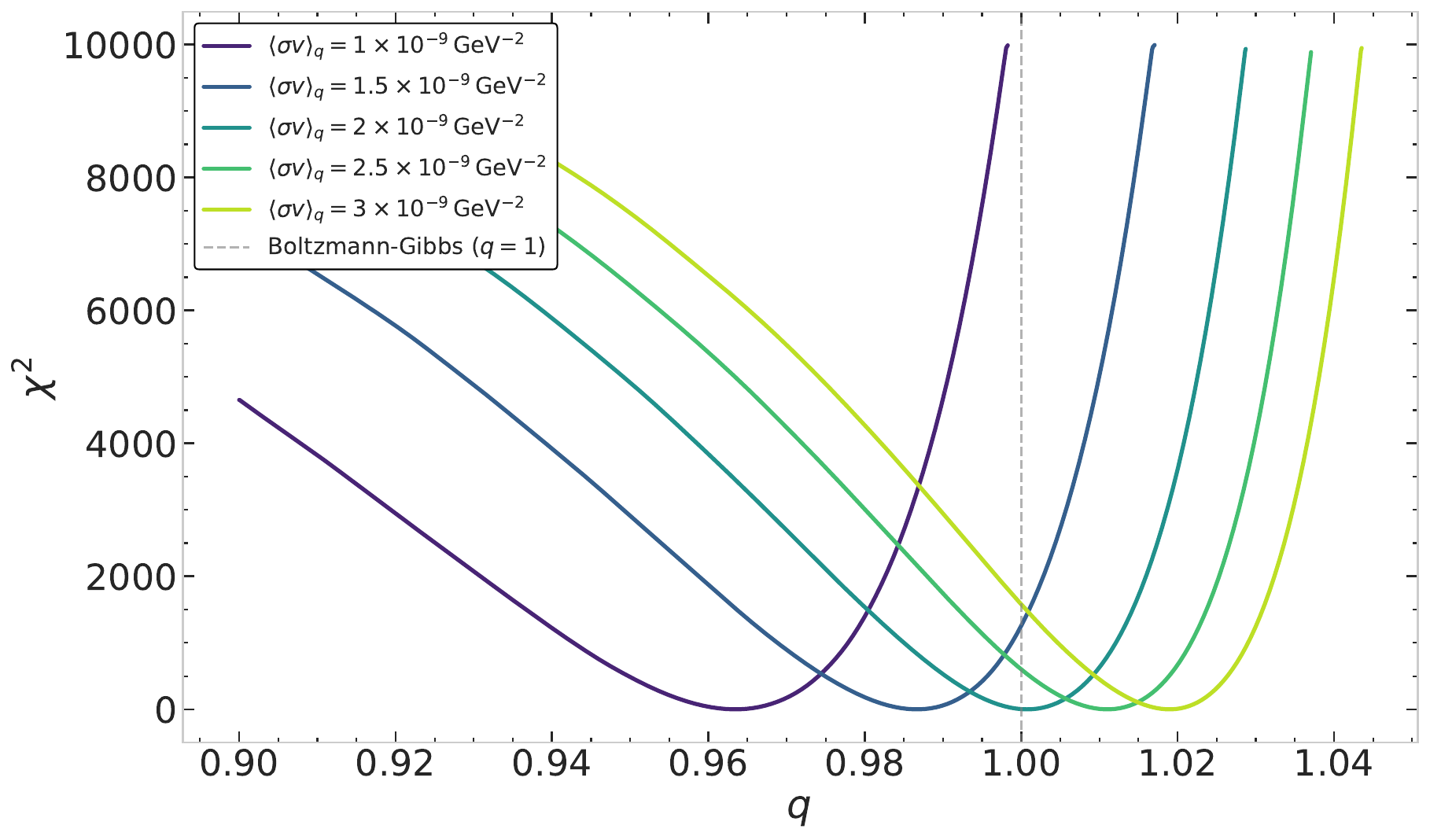}
    \caption{Absolute $\chi^2(q)$ profiles at fixed mass (cross section scan). For a representative $m_\chi = 100~\mathrm{GeV}$, curves are shown for different $s$-wave amplitudes $a\!\equiv\!\langle\sigma v\rangle_{s\text{-wave}}$ (in $\mathrm{GeV}^{-2}$). The smoothing highlights a valley of nearly equivalent $(q,a)$ solutions reproducing $\Omega_{c}h^2 = 0.120\pm0.001$, underscoring the role of $a$ as a nuisance parameter in a model-independent setup.}
    \label{fig:chisigma}
\end{figure}

\begin{figure}[!t]
    \centering
    \includegraphics[width=0.9\linewidth]{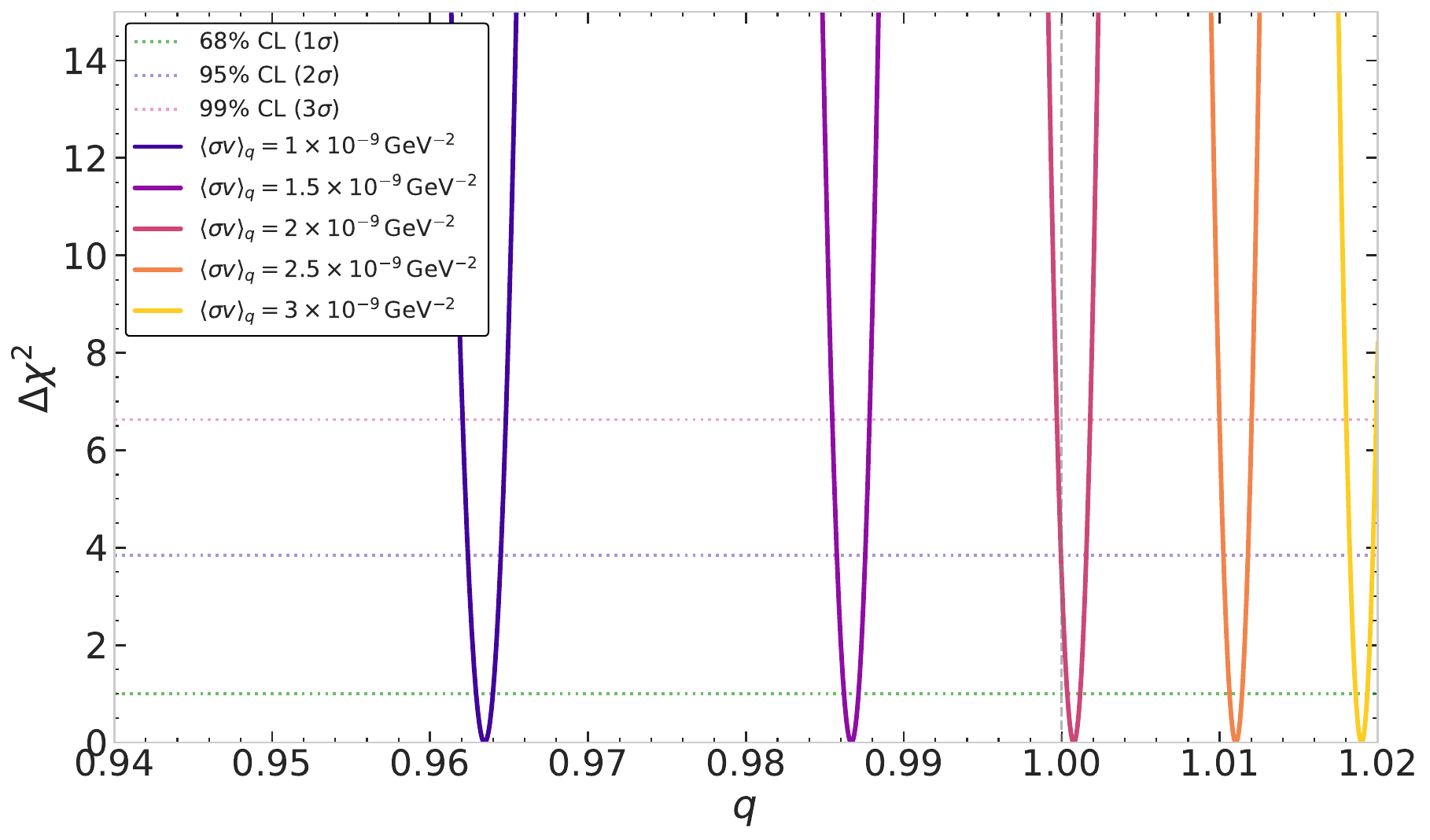}
    \caption{Profile likelihood $\Delta\chi^2(q)$ (cross section scan). The $\Delta\chi^2(q)$ curves corresponding to the previous panel are shown with 68\%, 95\%, and 99\% confidence-level lines, from which the confidence intervals in $q$ are read for each choice of $a$. The accumulation of compatible solutions effectively narrows the $q$ range around the best fit.}
    \label{fig:chisigmadelta}
\end{figure}

We then turn to the cross section scan at fixed mass (Figs.~\ref{fig:chisigma} and \ref{fig:chisigmadelta}). Here the minima are well separated for different $s$-wave amplitudes $a$, and modest changes in $a$ shift the preferred $q$ by amounts larger than the $68\%\text{-}95\%$ bands. This reveals a pronounced $(q,a)$ trade-off: in practice $a$ behaves as a nuisance parameter that traces an extended valley of nearly degenerate solutions reproducing $\Omega_{\chi}h^2$. We present the panels in this order to match the figure sequence and to reflect that current data do not yet fix $m_\chi$ or $\langle\sigma v\rangle_q$; a future measurement of either would break much of the $(q,a,m_\chi)$ degeneracy and sharpen the bounds on $q$.
\begin{figure}[!t]
    \centering
    \includegraphics[width=0.9\linewidth]{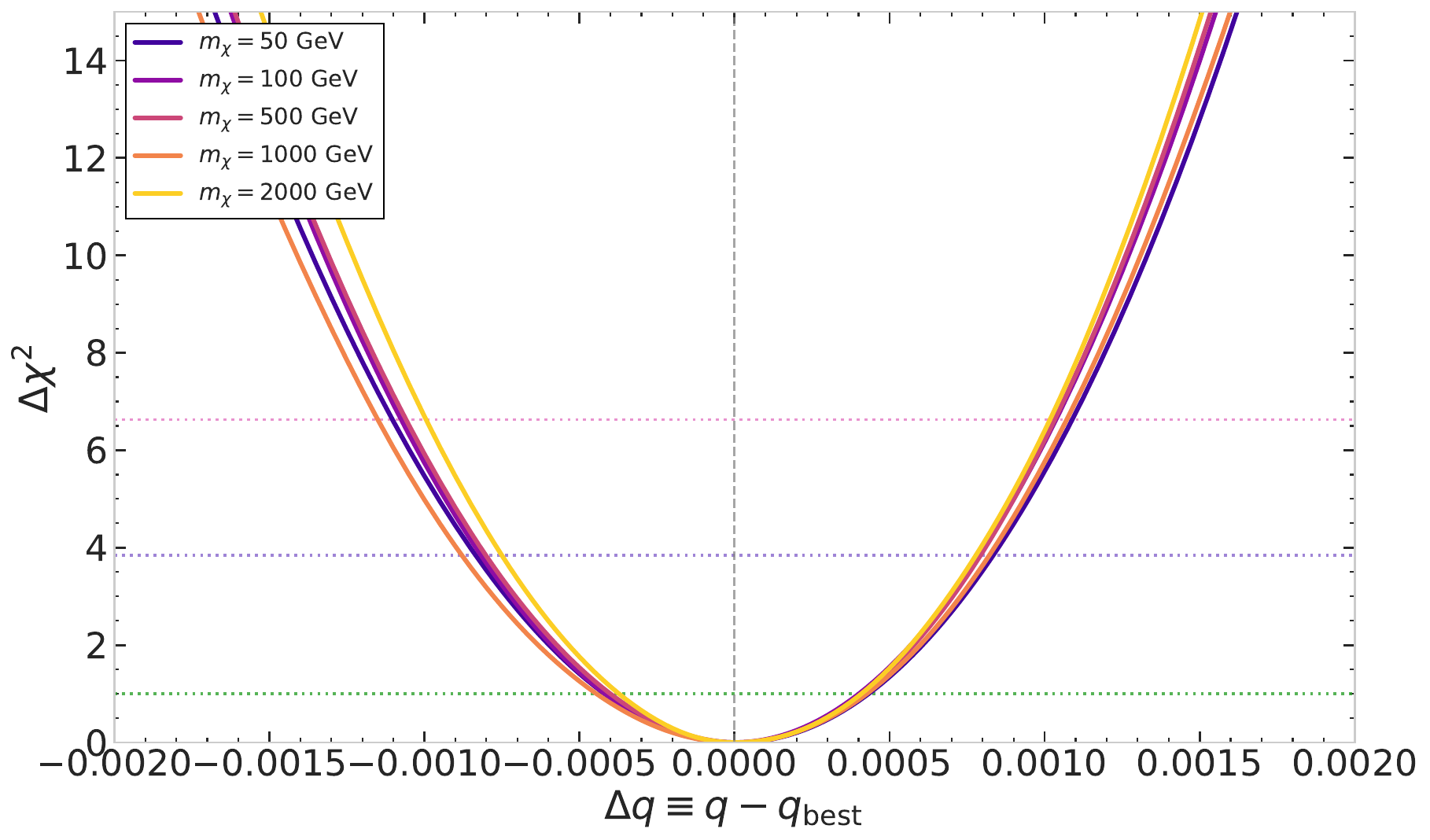}
\caption{Profile likelihood $\Delta\chi^2(q)$ (mass scan, centered). We show $\Delta\chi^2(q)\equiv \chi^2(q)-\chi^2_{\min}$ for $m_\chi=\{50,100,500,1000,2000\}\,\mathrm{GeV}$ with the horizontal axis centered at $q-q_{\rm best}$. Horizontal lines indicate the $68\%$, $95\%$, and $99\%$ confidence levels for one effective parameter. The clustering of minima across masses keeps the favored $q$ close to the extensive limit, consistent with the weak mass sensitivity discussed in the text.}
    \label{fig:centeredmasa}
\end{figure}
\begin{figure}[!t]
    \centering
    \includegraphics[width=0.9\linewidth]{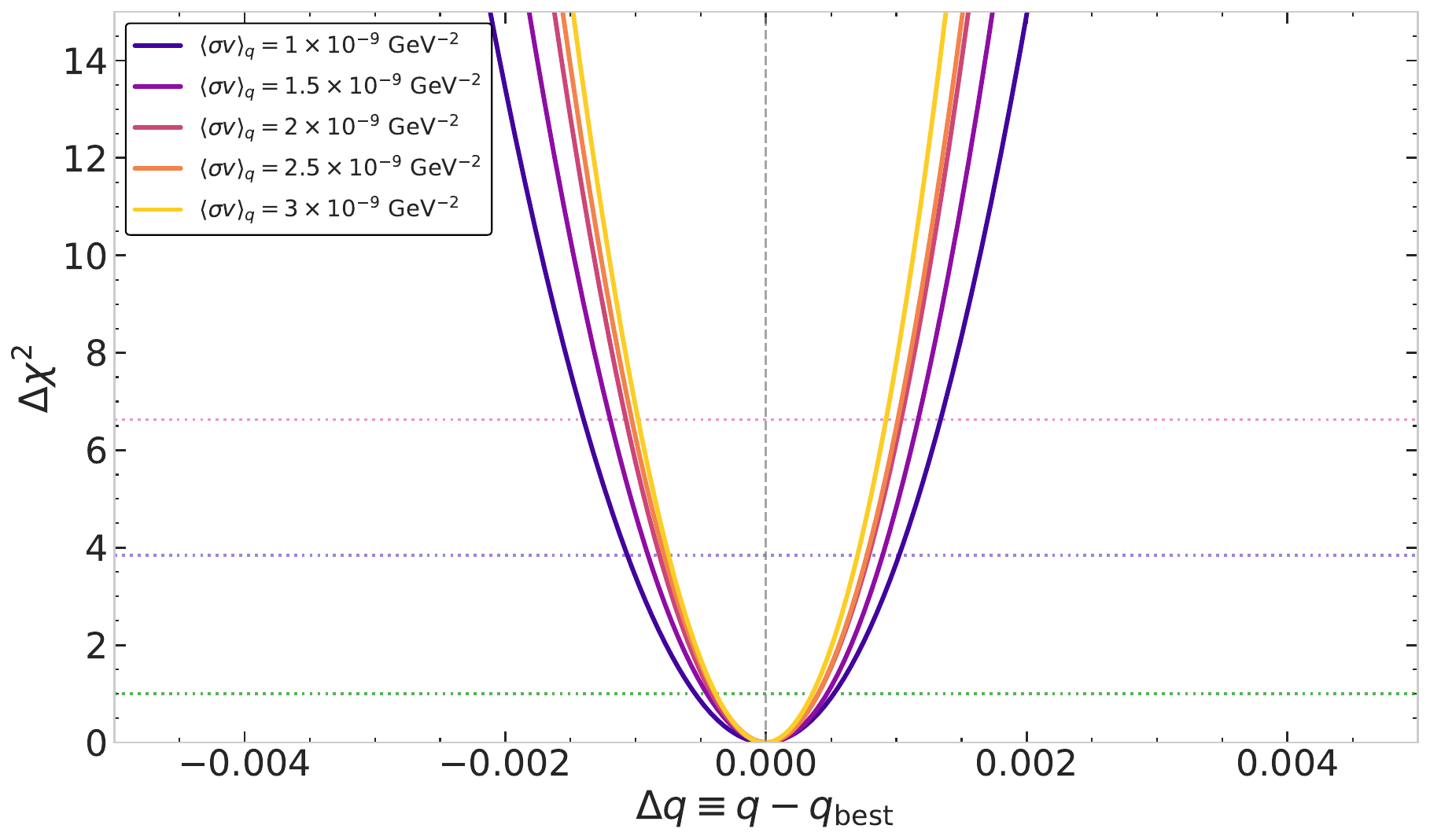}
\caption{Profile likelihood $\Delta\chi^2(q)$ (cross section scan, centered). For $m_\chi=100\,\mathrm{GeV}$, curves correspond to different $s$-wave amplitudes $a$ (legend shows $\langle\sigma v\rangle_q$ values in $\mathrm{GeV}^{-2}$); the horizontal axis is centered at $q-q_{\rm best}$. Modest changes in $a$ shift the preferred $q$ by amounts larger than the $68$-$95\%$ bands, tracing an extended $(q,a)$ valley of nearly degenerate solutions. Horizontal lines denote the $68\%$, $95\%$, and $99\%$ confidence levels.}
    \label{fig:centeredsigma}
\end{figure}

To compare widths independently of the minima locations, we recenter the profiles by $\Delta q \equiv q - q_{\rm best}$, where $q_{\rm best}$ is the per-curve best-fit value that reproduces $\Omega_{c}h^2 = 0.120\pm0.001$ (Figs.~\ref{fig:centeredmasa} and \ref{fig:centeredsigma}). In this representation all curves share a common origin at $\Delta q=0$, so differences in shape and width are no longer masked by shifts in the preferred $q$. The centered panels make explicit that the cross section scan yields broader $\Delta\chi^2(q)$ profiles than the mass scan: the curves obtained by varying $\langle\sigma v\rangle_q$ extend farther in $\Delta q$ while still lying within the $68$–$95\%$ confidence bands, whereas the mass-scan profiles remain comparatively narrow around their minima. In other words, the standard freeze-out hierarchy is preserved: variations of the thermally averaged cross section $\langle\sigma v\rangle_q$ are the primary lever modifying the relic abundance, efficiently moving the prediction across the Planck band, whereas changes in $m_\chi$ mainly shift the minimum without significantly broadening the profile. Thus, even in the nonextensive setup, the relic density is more sensitive to the annihilation strength than to the WIMP mass, and the degeneracy in $(q,a,m_\chi)$ is largely organized along directions dominated by $\langle\sigma v\rangle_q$ rather than by $m_\chi$.

\section{Conclusions}
\label{sec:conclusions}

In this work we developed a $q$-generalized framework for WIMPs in the freeze-out scenario by using nonextensive statistical mechanics. The construction combines exact $q$-exponential distributions for the early-universe plasma,  with ultra-relativistic background rescalings in the radiation map $R_{\rho}(q)$, and a $q$-generalized Boltzmann equation for the comoving abundance with a consistent treatment of the thermally averaged annihilation rate $\langle \sigma v\rangle_q$ via partial waves approximation. The guiding principle throughout was to avoid ad hoc deformations in the nonrelativistic particle sector at decoupling, computing instead the relevant moments directly from the exact phase-space integrals.

At the thermodynamic level, the ultra-relativistic mapping based on $R_{\rho}(q)$ provides a compact description of how nonextensivity reshapes $g_{*}(T)$, $g_{*s}(T)$, the entropy density, and the Hubble rate $H(T)$, thus propagating into the kinetic prefactor of the $q$-Boltzmann equation. The corresponding number-density map $R_{n}(q)$ was \textit{not} applied at freeze-out, since its derivation holds in the ultra-relativistic regime, while chemical decoupling takes place for nonrelativistic WIMPs. This choice preserves the consistency between the collision term and the exact $q$-integrals that define $n_{{\rm eq},q}$ and $\langle v_{\rm rel}^{2}\rangle_{q}$.

From the dynamical side, solving the $q$-Boltzmann equation shows that the freeze-out parameter $x_f$ increases monotonically with $q$ for fixed $m_\chi$. The trend arises from the interplay of the $q$-logarithmic inversion in the transcendental condition for $x_f$ and the background rescaling $H_q \propto \sqrt{R_\rho(q)}$. For $q>1$, power-law tails in $e_q$ enhance the velocity moments entering $\langle \sigma v\rangle_q \approx a + b \langle v_{\rm rel}^{2}\rangle_q$, delaying decoupling. For $q<1$, finite support suppresses those moments and advances decoupling. Step-like features induced by the Standard-Model thresholds in $g_{*}(T)$ and $g_{*s}(T)$ are inherited by $g_{*,q}$ and $g_{*s,q}$.
Confronting the predictions with the measured density parameter $\Omega_{c}h^{2}=0.120\pm0.001$ reveals a characteristic structure in $\chi^{2}(q)$. At fixed thermally averaged cross section, the minima across different $m_\chi$ cluster tightly, keeping the favored $q$ close to the extensive limit. In contrast, at fixed mass the preferred $q$ shifts noticeably with moderate changes in the $s$-wave amplitude $a$, exposing a pronounced $(q,a)$ trade-off: $a$ effectively behaves as a nuisance parameter tracing a valley of nearly degenerate solutions that reproduce the observed relic abundance. This hierarchy confirms that, within a model-independent setup, variations in $\langle \sigma v\rangle_q$ dominate the sensitivity to nonextensivity, while the WIMP mass mainly induces a mild relocation of the best-fit region.

The analysis clarifies the scope and boundaries of the approach. First, the use of $R_{\rho}(q)$ is rigorously justified in the ultra-relativistic sector, where its integral definition converges up to the known bounds in $q$; its impact at freeze-out enters only through the background (expansion and entropy) and not through an explicit rescaling of nonrelativistic number densities. Second, the treatment of $\langle \sigma v\rangle_q$ via exact $q$-moments captures, by construction, both the enhancement of high-energy tails for $q>1$ and the compact support for $q<1$. Third, the statistical interpretation makes explicit the partial degeneracies among $(q,a,m_\chi)$ when only $\Omega_{c}h^2$ is used as constraint.

In summary, the $q$-generalized freeze-out framework developed here establishes a consistent and tractable bridge between nonextensive statistical mechanics and WIMP cosmology. It identifies robust qualitative signatures (monotonic $x_f(q)$, hierarchy of sensitivities, background-imprinted kinks) and quantifies where present data place the strongest leverage on $q$.

\section*{Acknowledgements}

MPG, and RAL. acknowledge Vicerrectoría de Investigación y Desarrollo Tecnológico (VRIDT) at Universidad Católica del Norte (UCN) for the scientific support provided by Núcleo de Investigación en Simetrías y la Estructura del Universo (NISEU-UCN), Resolución VRIDT N°200/2025.

MPG. acknowledges to my fellow roommates of the graduate program at \textit{Universidad Católica del Norte}.

MPG. acknowledges the financial support of the \textit{Dirección general de postgrado}.

\appendix
\section{$q$-exponential and $q$-logarithm}
\label{anexo:1}
\begin{figure}[!h]
    \centering
    \includegraphics[width=1\linewidth]{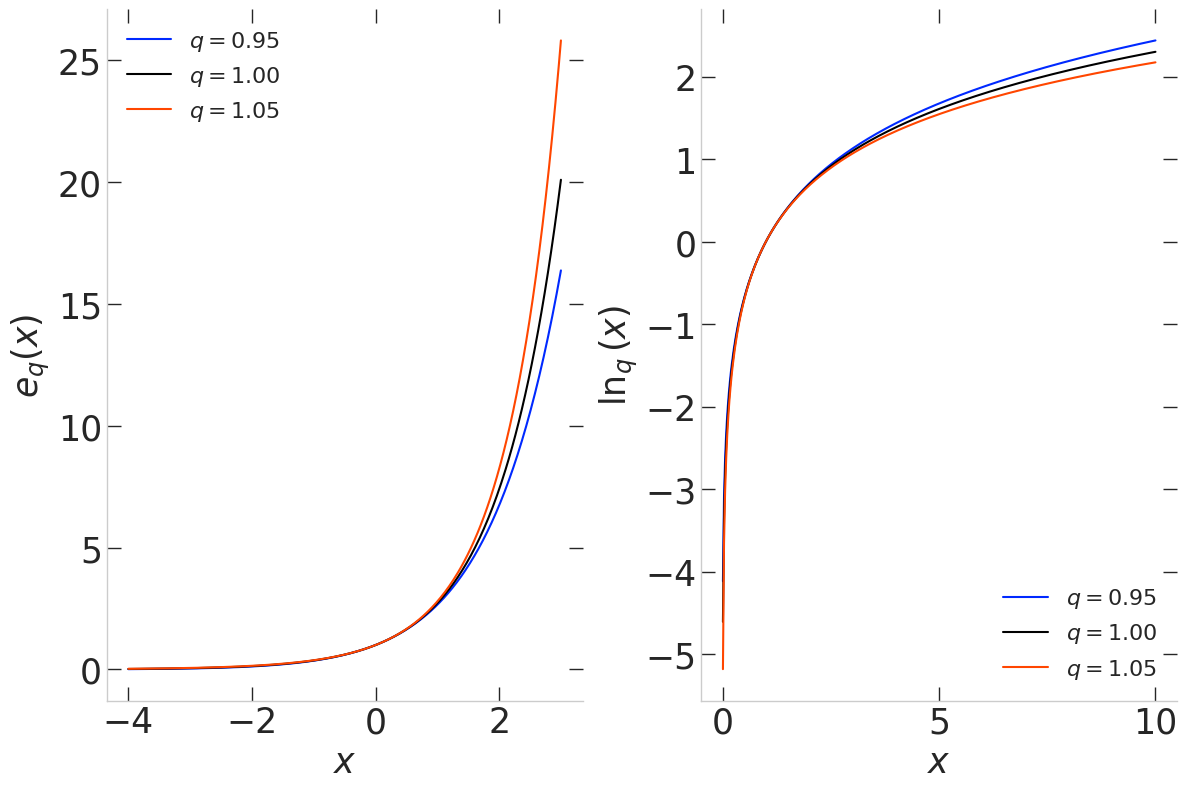}
    \caption{$q$-exponential $e_q(x)$ (left) and $q$-logarithm $\ln_q(x)$ (right) for $q \in \{0.95,\,1.00,\,1.05\}$. For $x>0$ the $q>1$ curve lies above the standard case, enhancing the growth of $e_q(x)$, while $q<1$ suppresses it; for $\ln_q(x)$ the deformation is such that $q>1$ compresses and $q<1$ stretches the curve with respect to $\ln x$. In both panels all curves meet at $e_q(0)=1$ and $\ln_q(1)=0$ and smoothly approach the usual exponential and logarithm in the $q\to 1$ limit.}
    \label{fig:placeholder}
\end{figure}
Nonextensive statistical mechanics is formulated in terms of the $q$-exponential and the $q$-logarithm introduced in Eq.~\eqref{eq:qexp}, which implement a controlled, one-parameter deformation of the standard exponential and logarithmic functions and smoothly reduce to them in the extensive limit $q\to 1$~\cite{Tsallis2009Book}. Within our framework, $e_q(x)$ sets the statistical weights that enter the $q$-generalized distribution functions, while $\ln_q(x)$ appears in the construction of thermodynamic quantities and in the inversion of transcendental relations such as the freeze-out condition. Departures $q\neq 1$ thus encode modified tails, effective support, and additivity properties directly at the level of these generalized functions, propagating to all macroscopic observables built from them. The qualitative impact of varying $q$ on both $e_q(x)$ and $\ln_q(x)$ for representative values is illustrated in Fig.~\ref{fig:placeholder}.

\section{Convergence of integrals}
\label{anexo:2}

The $q$-rescaled observables UR used in the main text can be written in terms of
\begin{equation}
  I_m(q) \;\equiv\; \int_0^{z_{\max}(q)} z^m\, e_q(-z)\,dz,
  \label{eq:Imq-app}
\end{equation}
with $m\ge 0$ and $z\equiv p/T$. The $q$-exponential is given by Eq.~\eqref{eq:qexp}, and its support reads
\begin{equation}
  z_{\max}(q) \;=\;
  \begin{cases}
    \dfrac{1}{1-q}, & q<1,\\[4pt]
    \infty, & q\ge 1.
  \end{cases}
  \label{eq:zmax-app-short}
\end{equation}
For $q<1$ the integration domain is finite and $z^m e_q(-z)$ is continuous on $[0,z_{\max}(q)]$, so all moments $I_m(q)$ converge for any $m\ge 0$.

For $q>1$ the support extends to infinity and convergence is controlled by the large-$z$ tail. Using
\begin{equation}
  e_q(-z)
  = \bigl[\,1 + (q-1)z\,\bigr]^{-\frac{1}{q-1}}
  \sim z^{-\frac{1}{q-1}}
  \qquad (z\to\infty),\nonumber
\end{equation}
the integrand behaves as
\begin{equation}
  z^m e_q(-z) \;\sim\; z^{\,m - \frac{1}{q-1}}.
\end{equation}
The integral $\int^\infty dz\,z^\alpha$ converges at the upper limit only if $\alpha<-1$, so we require
\begin{equation}
  m - \frac{1}{q-1} < -1
  \quad\Longrightarrow\quad
  q < 1 + \frac{1}{m+1}.
  \label{eq:qbound-general-short}
\end{equation}
\medskip
In the UR Maxwell-Boltzmann limit, $n_q$ and $\rho_q$ correspond to $m=2$ and $m=3$, respectively:
\begin{align}
  n_q &\propto I_2(q)
  &&\Rightarrow&
  q &< 1 + \frac{1}{2+1} = \frac{4}{3},\\[4pt]
  \rho_q &\propto I_3(q)
  &&\Rightarrow&
  q &< 1 + \frac{1}{3+1} = \frac{5}{4}.
\end{align}
Thus the $q$-rescaled number and energy densities $R_n(q)~\eqref{eq:Rn-exacto}$ and $R_\rho(q)$~\eqref{eq:def-Rrho} are well defined for
\begin{equation}
  R_n(q):\; q<\frac{4}{3},
  \qquad
  R_\rho(q):\; q<\frac{5}{4},
\end{equation}
with automatic convergence for all $q<1$ due to compact support.


 \bibliographystyle{elsarticle-num} 
 \bibliography{apssamp}

@article{Tsallis1988,
  author    = {Constantino Tsallis},
  title     = {Possible generalization of Boltzmann-Gibbs statistics},
  journal   = {Journal of Statistical Physics},
  volume    = {52},
  number    = {1-2},
  pages     = {479--487},
  year      = {1988},
  doi       = {10.1007/BF01016429}
}

@article{CuradoTsallis1991,
  author    = {E. M. F. Curado and C. Tsallis},
  title     = {Generalized statistical mechanics: connection with thermodynamics},
  journal   = {Journal of Physics A: Mathematical and General},
  volume    = {24},
  pages     = {L69--L72},
  year      = {1991},
  doi       = {10.1088/0305-4470/24/2/009}
}

@article{TsallisMendesPlastino1998,
  author    = {C. Tsallis and R. S. Mendes and A. R. Plastino},
  title     = {The role of constraints within generalized nonextensive statistics},
  journal   = {Physica A},
  volume    = {261},
  pages     = {534--554},
  year      = {1998},
  doi       = {10.1016/S0378-4371(98)00437-3}
}

@book{Tsallis2009Book,
  author    = {Constantino Tsallis},
  title     = {Introduction to Nonextensive Statistical Mechanics},
  publisher = {Springer},
  address   = {New York},
  year      = {2009},
  doi       = {10.1007/978-0-387-85359-8}
}

@article{Buyukilic1995,
  author    = {Fevzi B{\"u}y{\"u}kili{\c{c}} and Dogan Demirhan and Atilla G{\"u}le{\c{c}}},
  title     = {A statistical mechanical approach to generalized statistics of quantum and classical gases},
  journal   = {Physics Letters A},
  volume    = {197},
  number    = {3},
  pages     = {209--220},
  year      = {1995},
  doi       = {10.1016/0375-9601(94)00982-6}
}

@inproceedings{FRIEMAN_1994,
   title={THE STANDARD COSMOLOGY},
   url={http://dx.doi.org/10.1142/9789814503785_0012},
   DOI={10.1142/9789814503785_0012},
   booktitle={The Building Blocks of Creation},
   publisher={WORLD SCIENTIFIC},
   author={FRIEMAN, JOSHUA A.},
   year={1994},
   month=oct, pages={421–468} }

@book{KolbTurner1990,
  author = {Kolb, Edward W. and Turner, Michael S.},
  title = {The Early Universe},
  publisher = {Addison-Wesley},
  year = {1990}
}

@article{GondoloGelmini1991,
  author = {Gondolo, Paolo and Gelmini, Graciela},
  title = {Cosmic abundances of stable particles: Improved analysis},
  journal = {Nucl. Phys. B},
  volume = {360},
  pages  = {145--179},
  year   = {1991},
  doi    = {10.1016/0550-3213(91)90438-4}
}

@article{2020,
   title={Planck2018 results: VI. Cosmological parameters},
   volume={641},
   ISSN={1432-0746},
   url={http://dx.doi.org/10.1051/0004-6361/201833910},
   DOI={10.1051/0004-6361/201833910},
   journal={Astronomy \& Astrophysics},
   publisher={EDP Sciences},
   author={Aghanim, N. and Akrami, Y. and Ashdown, M. and Aumont, J. and Baccigalupi, C. and Ballardini, M. and Banday, A. J. and Barreiro, R. B. and Bartolo, N. and Basak, S. and Battye, R. and Benabed, K. and Bernard, J.-P. and Bersanelli, M. and Bielewicz, P. and Bock, J. J. and Bond, J. R. and Borrill, J. and Bouchet, F. R. and Boulanger, F. and Bucher, M. and Burigana, C. and Butler, R. C. and Calabrese, E. and Cardoso, J.-F. and Carron, J. and Challinor, A. and Chiang, H. C. and Chluba, J. and Colombo, L. P. L. and Combet, C. and Contreras, D. and Crill, B. P. and Cuttaia, F. and de Bernardis, P. and de Zotti, G. and Delabrouille, J. and Delouis, J.-M. and Di Valentino, E. and Diego, J. M. and Doré, O. and Douspis, M. and Ducout, A. and Dupac, X. and Dusini, S. and Efstathiou, G. and Elsner, F. and Enßlin, T. A. and Eriksen, H. K. and Fantaye, Y. and Farhang, M. and Fergusson, J. and Fernandez-Cobos, R. and Finelli, F. and Forastieri, F. and Frailis, M. and Fraisse, A. A. and Franceschi, E. and Frolov, A. and Galeotta, S. and Galli, S. and Ganga, K. and Génova-Santos, R. T. and Gerbino, M. and Ghosh, T. and González-Nuevo, J. and Górski, K. M. and Gratton, S. and Gruppuso, A. and Gudmundsson, J. E. and Hamann, J. and Handley, W. and Hansen, F. K. and Herranz, D. and Hildebrandt, S. R. and Hivon, E. and Huang, Z. and Jaffe, A. H. and Jones, W. C. and Karakci, A. and Keihänen, E. and Keskitalo, R. and Kiiveri, K. and Kim, J. and Kisner, T. S. and Knox, L. and Krachmalnicoff, N. and Kunz, M. and Kurki-Suonio, H. and Lagache, G. and Lamarre, J.-M. and Lasenby, A. and Lattanzi, M. and Lawrence, C. R. and Le Jeune, M. and Lemos, P. and Lesgourgues, J. and Levrier, F. and Lewis, A. and Liguori, M. and Lilje, P. B. and Lilley, M. and Lindholm, V. and López-Caniego, M. and Lubin, P. M. and Ma, Y.-Z. and Macías-Pérez, J. F. and Maggio, G. and Maino, D. and Mandolesi, N. and Mangilli, A. and Marcos-Caballero, A. and Maris, M. and Martin, P. G. and Martinelli, M. and Martínez-González, E. and Matarrese, S. and Mauri, N. and McEwen, J. D. and Meinhold, P. R. and Melchiorri, A. and Mennella, A. and Migliaccio, M. and Millea, M. and Mitra, S. and Miville-Deschênes, M.-A. and Molinari, D. and Montier, L. and Morgante, G. and Moss, A. and Natoli, P. and Nørgaard-Nielsen, H. U. and Pagano, L. and Paoletti, D. and Partridge, B. and Patanchon, G. and Peiris, H. V. and Perrotta, F. and Pettorino, V. and Piacentini, F. and Polastri, L. and Polenta, G. and Puget, J.-L. and Rachen, J. P. and Reinecke, M. and Remazeilles, M. and Renzi, A. and Rocha, G. and Rosset, C. and Roudier, G. and Rubiño-Martín, J. A. and Ruiz-Granados, B. and Salvati, L. and Sandri, M. and Savelainen, M. and Scott, D. and Shellard, E. P. S. and Sirignano, C. and Sirri, G. and Spencer, L. D. and Sunyaev, R. and Suur-Uski, A.-S. and Tauber, J. A. and Tavagnacco, D. and Tenti, M. and Toffolatti, L. and Tomasi, M. and Trombetti, T. and Valenziano, L. and Valiviita, J. and Van Tent, B. and Vibert, L. and Vielva, P. and Villa, F. and Vittorio, N. and Wandelt, B. D. and Wehus, I. K. and White, M. and White, S. D. M. and Zacchei, A. and Zonca, A.},
   year={2020},
   month=sep, pages={A6}
}

@article{Pessah_2001,
   title={Statistical mechanics and the description of the early universe. (I). Foundations for a slightly non-extensive cosmology},
   volume={297},
   ISSN={0378-4371},
   url={http://dx.doi.org/10.1016/S0378-4371(01)00235-7},
   DOI={10.1016/s0378-4371(01)00235-7},
   number={1–2},
   journal={Physica A: Statistical Mechanics and its Applications},
   publisher={Elsevier BV},
   author={Pessah, M.E. and Torres, Diego F. and Vucetich, H.},
   year={2001},
   month=aug, pages={164–200} }

@misc{PDG2024_AstroConsts,
  author       = {Particle Data Group},
  title        = {Astrophysical Constants \& Parameters (rev.)},
  howpublished = {in \emph{Review of Particle Physics} (2024)},
  year         = {2024},
  note         = {Table of astrophysical constants and cosmological parameters},
  url          = {https://pdg.lbl.gov/2024/reviews/rpp2024-rev-astrophysical-constants.pdf}
}

@article{Saikawa_2020,
   title={Precise WIMP dark matter abundance and Standard Model thermodynamics},
   volume={2020},
   ISSN={1475-7516},
   url={http://dx.doi.org/10.1088/1475-7516/2020/08/011},
   DOI={10.1088/1475-7516/2020/08/011},
   number={08},
   journal={Journal of Cosmology and Astroparticle Physics},
   publisher={IOP Publishing},
   author={Saikawa, Ken’ichi and Shirai, Satoshi},
   year={2020},
   month=aug, pages={011–011} }

@article{Hindmarsh_2005,
   title={Dark matter of weakly interacting massive particles and the QCD equation of state},
   volume={71},
   ISSN={1550-2368},
   url={http://dx.doi.org/10.1103/PhysRevD.71.087302},
   DOI={10.1103/physrevd.71.087302},
   number={8},
   journal={Physical Review D},
   publisher={American Physical Society (APS)},
   author={Hindmarsh, Mark and Philipsen, Owe},
   year={2005},
   month=apr }

@article{Scherrer:1986,
  author  = {R. J. Scherrer and M. S. Turner},
  title   = {On the relic, cosmic abundance of stable, weakly interacting massive particles},
  journal = {Phys. Rev. D},
  year    = {1986},
  volume  = {33},
  pages   = {1585}
}

@article{LaineSchroder:2006,
  author  = {M. Laine and Y. Schr{\"o}der},
  title   = {Quark mass thresholds in QCD thermodynamics},
  journal = {Phys. Rev. D},
  year    = {2006},
  volume  = {73},
  pages   = {085009}
}

@article{GriestSeckel:1991,
  author  = {K. Griest and D. Seckel},
  title   = {Three exceptions in the calculation of relic abundances},
  journal = {Phys. Rev. D},
  year    = {1991},
  volume  = {43},
  pages   = {3191}
}

@article{LimaPlastino:2001,
  author  = {J. A. S. Lima and R. Silva and A. R. Plastino},
  title   = {Nonextensive Thermostatistics and the $H$ Theorem},
  journal = {Phys. Rev. Lett.},
  year    = {2001},
  volume  = {86},
  pages   = {2938}
}

@misc{rueter2020darkmatterfreezetsallis,
      title={Dark Matter Freeze Out with Tsallis Statistics in the Early Universe}, 
      author={Thomas D. Rueter and Thomas G. Rizzo and JoAnne L. Hewett},
      year={2020},
      eprint={1911.11254},
      archivePrefix={arXiv},
      primaryClass={hep-ph},
      url={https://arxiv.org/abs/1911.11254}, 
}

@Article{e25111495,
AUTHOR = {Jizba, Petr and Lambiase, Gaetano},
TITLE = {Constraints on Tsallis Cosmology from Big Bang Nucleosynthesis and the Relic Abundance of Cold Dark Matter Particles},
JOURNAL = {Entropy},
VOLUME = {25},
YEAR = {2023},
NUMBER = {11},
ARTICLE-NUMBER = {1495},
URL = {https://www.mdpi.com/1099-4300/25/11/1495},
PubMedID = {37998187},
ISSN = {1099-4300},
DOI = {10.3390/e25111495}
}

@article{Arcadi:2017kky,
    author = "Arcadi, Giorgio and Dutra, Ma{\'\i}ra and Ghosh, Pradipta and Lindner, Manfred and Mambrini, Yann and Pierre, Mathias and Profumo, Stefano and Queiroz, Farinaldo S.",
    title = "{The waning of the WIMP? A review of models, searches, and constraints}",
    eprint = "1703.07364",
    archivePrefix = "arXiv",
    primaryClass = "hep-ph",
    doi = "10.1140/epjc/s10052-018-5662-y",
    journal = "Eur. Phys. J. C",
    volume = "78",
    number = "3",
    pages = "203",
    year = "2018"
}



\end{document}